\documentclass[10pt,a4paper]{article}
\usepackage{amsmath}
\usepackage{graphicx}
\usepackage{enumerate}
\usepackage{amssymb}
\usepackage{amsthm}
\usepackage{natbib}
\usepackage{fullpage}
\usepackage{mathtools}
\usepackage{url} 
\usepackage[dvipsnames,svgnames,x11names]{xcolor}
\usepackage[colorlinks = true, linkcolor = red!50!black,
citecolor = ProcessBlue!50!RoyalBlue, urlcolor = green!50!black,
filecolor=Maroon]{hyperref}
\usepackage{datetime2}

\usepackage{multirow}
\usepackage{booktabs}
\usepackage{dsfont}
\usepackage{lscape}
\usepackage[labelfont=bf, font=small]{caption}

\newcommand{\RR}{\mathbb{R}}

\newcommand{\indep}{\protect{\perp\!\!\!\perp}}
\newcommand{\given}{\mid}
\newcommand{\1}{\mathds{1}}
\newcommand{\pkg}[1]{\textbf{#1}}
\newcommand{\proglang}[1]{\textsf{#1}}
\DeclareMathOperator{\Ex}{\mathbb{E}}
\DeclareMathOperator{\Prob}{\mathbb{P}}

\theoremstyle{plain}

\theoremstyle{plain}
\newtheorem{proposition}{Proposition}
\AtBeginDocument{}

\renewcommand{\hat}{\widehat}

\title{\bf Rethinking player evaluation in sports:\\Goals above expectation and beyond}
\author{Robert Bajons\footnote{Corresponding Author:
  \href{mailto:robert.bajons@wu.ac.at}{\texttt{robert.bajons@wu.ac.at}}
  } \and Lucas Kook}
\date{\scriptsize 
Institute for Statistics and Mathematics, 
Vienna University of Business and Economics, 
Vienna, Austria}

\begin{document}

\maketitle

\begin{abstract}
A popular quantitative approach to evaluating player performance in sports
involves comparing an observed outcome to the expected outcome ignoring player
involvement, which is estimated using statistical or machine learning methods.
In soccer, for instance, goals above expectation (GAX) of a player measure how
often shots of this player led to a goal compared to the model-derived expected
outcome of the shots. Typically, sports data analysts rely on flexible machine
learning models, which are capable of handling complex nonlinear effects and
feature interactions, but fail to provide valid statistical inference due to
finite-sample bias and slow convergence rates. In this paper, we close this gap
by presenting a framework for player evaluation with metrics derived from
differences in actual and expected outcomes using flexible machine learning
algorithms, which nonetheless allows for valid frequentist inference. We first
show that the commonly used metrics are directly related to Rao's score test in
parametric regression models for the expected outcome. Motivated by this finding
and recent developments in double machine learning, we then propose the use of
residualized versions of the original metrics. For GAX, the residualization step
corresponds to an additional regression predicting whether a given player would
take the shot under the circumstances described by the features. We further
relate metrics in the proposed framework to player-specific effect estimates in
interpretable semiparametric regression models, allowing us to infer directional
effects, e.g., to determine players that have a positive impact on the outcome.
Our primary use case are GAX in soccer. We further apply our framework to
evaluate goal-stopping ability of goalkeepers, shooting skill in basketball,
quarterback passing skill in American football, and injury-proneness of soccer
players.
\end{abstract}

\section{Introduction}
\label{sec:Intro}

The availability of novel and granular data has vastly transformed the way
professional sport is analyzed. The field of sports analytics, a research area
combining statistical and machine learning and sports science, has attracted a
lot of interest \citep{Lopez18randomness}, and the insights generated from
analyzing data with statistical tools are directly affecting the dynamics of
games in various sports \citep{Baumer2023BigIdeas}. An area where sports
analytics plays a key role is the recruitment of players. To efficiently assess
and detect undervalued players, it is of fundamental importance to accurately
measure a player's skills. In dynamic games such as soccer, American football,
ice hockey, or basketball, a quantitative approach to evaluate player
performance relies on estimating an expected value for an outcome based on
contextual features describing the game state
\citep{Davis2024Evaluation,Brill2024EP}. The observed outcome can then
be compared with the expected outcome to determine whether a player's
performance deviates from the model’s expectation.
Mathematically, we can
express the value of a player as \begin{equation}\label{eq:player_value}
\sum_{j=1}^{N} (Y_j - \hat h(Z_j))X^p_j. \end{equation} Here, $N$ denotes the
number of observations of interest for evaluating a
player, e.g.~shots or more general any actions (passes, dribbles, or crosses) in
soccer, ice hockey, or basketball; $X^p$ is an indicator for player
$p$ participating;
$\hat h$ is an estimator for  $h(Z) \coloneqq \Ex[Y \given Z]$,
the conditional expectation of the 
outcome for a game state represented by $Z$ (excluding player participation);
$Y$ is the outcome of interest, such as (field) goals
in soccer, ice hockey, or basketball. In Table~\ref{tab:applications}, we
highlight a number of player evaluation metrics that can be expressed as in
\eqref{eq:player_value}. While this list makes no attempt at being exhaustive,
these are the use cases considered empirically in this work.

\begin{table}[!ht]
\centering
  \caption{Player evaluation metrics that fall within our framework and for which we present
  applications. Our framework extends all
  metrics by an additional residualization 
  step, taking into account whether a given player would undertake an action, such
  as a shot, under circumstances described by the same features used in the outcome
  regression. Details on the residualization step are given in Section~\ref{sec:PLLM}.
  }
  \label{tab:applications}
    \resizebox{\linewidth}{!}{
    \begin{tabular}{lrrr}
      \toprule
      \bf Discipline & \bf Metric & \bf Outcome & \bf Section \\
      \midrule
       Soccer & Goals above expectation
       (GAX)  & Goal (binary) & Section~\ref{subsec:GAX_vs_rGAX} \\
      Soccer &  Goals saved above expectation (GSAX)  & Goal saved (binary) &
      Section~\ref{subsec:GSAX_vs_rGSAX} \\
      Basketball &  Quantified shooter impact (qSI)  & Field goals (binary or score) &
      Appendix~\ref{subsec:SQ_SI}\\
      American football &  Completion percentage above expectation (CPAE)  & Pass completion (binary) &
      Appendix~\ref{subsec:CPAE}\\
      Soccer &  Injuries above expectation (IAX) & Time to first injury
      (right-censored) & Appendix~\ref{subsec:TTI}\\
      \bottomrule
    \end{tabular}
    }
\end{table}

In this work, we primarily focus on goals above expectation (GAX) in soccer as a
means for analyzing shooting skills. In order to quantify the shooting skills of
players, a crucial first step is to identify adequate means for analyzing
shooting ability. Since goals are the most important outcomes in soccer, a
classical strategy is to count the number of goals scored by a player. However,
there are two fundamental issues with this approach: First, goals are
exceptionally rare, with an average of two to three goals per match
\citep{Scarf21skillandchance}. Second, as many authors have pointed out
\citep[e.g.,][]{AB21,HK23}, scoring a goal largely depends on the circumstances of a shot.
A shot closer to the goal with only the goalkeeper in the way is far more likely
to result in a goal than a shot further away with a large number of defenders in
front of the shooter. These issues have been widely acknowledged in the sports
analytics literature and have led to the development of so-called expected goals
(xG) models. xG models assign a probability of success to each shot, taking into
account factors that influence the likelihood of scoring a goal from a shot. In
fact, this idea is not new, and an early version of an xG model has already been
proposed by \citet{PR97}. \citeauthor{PR97} used a logistic regression model for
the binary outcome of a shot and found that the most important factors for
successful shots were the shot location, the angle between the shot and the two
goalposts (henceforth, goalangle), and the body part with which the shot was
carried out (foot or head). Recently, access to novel data types such as the event
stream data (as described in Section~\ref{sec:data}) has led to rapid
development of xG models. Modern approaches are based on flexible machine
learning algorithms, such as extreme gradient boosting machines, that account
for non-linear and interaction effects, taking into account a detailed set of
shot-specific features \citep{RD20,AB21,HK23}. Therefore, xG models provide a
contextualized version of shots and goals, making them a popular tool for
analysis of teams and players. Furthermore, they are a main building block for
holistic approaches to model soccer games, such as expected possession value
models \citep{BFC21}. As a measure of a shot’s success, xG also serves as a
building block for evaluating shooting skill. In particular, we can use xG to
determine a player's GAX, defined as the summed differences between the actual
outcome of all shots taken by this player (1: goal, or 0: no goal) and the
probability of the respective shots to end in a goal as computed by the xG model
\citep{DR24}.

Using GAX as our primary use case has various reasons. First, being
based on xG, it is a very intuitive metric and easy to explain, hence attractive
for sports data analysts. Second, GAX recently received various forms of
criticism in scientific work. Particularly, GAX has been criticized for being
unstable over seasons, i.e., a player's GAX in one season is poorly predictive
of the player's GAX in the next season, for being prone to biases in the data,
and for not allowing for uncertainty quantification. This has resulted in GAX
being labeled a poor metric for evaluating shooting skills
\citep{DR23xmskill,BSPC24,DR24}.
While some of these criticisms, such as the small effective sample size innate
to soccer, cannot directly be remedied by methodological advancements, we
believe that the lack of uncertainty quantification and replicability can be
addressed by approaching the problem of player evaluation with modern machine
learning and statistical modeling techniques, namely double machine learning
\citep{chern2017double} and nonparametric conditional independence testing
\citep{SP20}. Throughout this paper, GAX will serve
as a key example for common pitfalls when trying to evaluate players
or key skills of players, such as their shooting ability, as we believe
the criticism of GAX largely translates to other sports.

\subsection{Our contributions}

The main contribution of this work is to introduce a general
framework for player evaluation by extending the idea
in Equation~\eqref{eq:player_value} to semiparametric
models. The starting point of our framework is the observation
that metrics of the form in~\eqref{eq:player_value} resemble score statistics
\citep{Rao48}, which allow for valid statistical inference when assuming a
parametric model. However, parametric models are restrictive and are often not
considered appropriate for modeling the complex nature of sports, as discussed
above. Therefore, popular approaches for evaluating players are
typically developed using modern machine learning tools, which do not directly
allow valid frequentist uncertainty quantification. Our proposed framework
closes this gap by casting player strengths as parameters in (generalized)
partially linear models and relating tests of player strength in these models to
nonparametric conditional independence tests, in particular, the
well-established Generalised Covariance Measure (GCM) test \citep{SP20}. As
such, the framework is also related to recent advancements in semiparametric
statistics \citep{kennedy2024semiparametric}, double machine learning
\citep{chern2017double}, and assumption-lean inference \citep{VD22ALR}. In
particular, we make the following contributions:

\begin{itemize}
    \item We propose a framework that allows for valid frequentist inference
    on player effects in the form of (directional) hypothesis tests
    (Proposition~\ref{thm:prp1}), even when using machine learning models for
    modeling relationships between outcome and features;
    \item We show that models within our framework relate to well-known
    semiparametric (generalized) partially linear models, which enable easy
    interpretation of player effects (Section~\ref{sec:PLLM});
     \item We
     apply the proposed framework primarily to the case
     of GAX
    and present various related approaches in different sports throughout
    the main text and appendix (Section~\ref{subsec:GAX_vs_rGAX}, see also
    Table~\ref{tab:applications}); 
    \item The proposed framework naturally provides a residualized version of
    GAX (which we call rGAX), which addresses aforementioned
    existing issues with using GAX as a measure to evaluate
    shooting skills of soccer players \citep{DR23xmskill,BSPC24,DR24}.
\end{itemize}

The rest of this paper is structured as follows. 
We begin Section~\ref{sec:problem-and-data} by describing the available
data structure.
In Section~\ref{sec:GAX_BG}, we then
recap standard approaches for deriving GAX and relate them to a player-specific
strength estimate in a classical parametric model. Section~\ref{sec:PLLM}
extends the ideas from Section~\ref{sec:GAX_BG} to a more flexible
semiparametric model. We present residualized GAX (rGAX) as an alternative to
GAX, and connect rGAX to a strength estimate in the semiparametric model.
In Section~\ref{sec:res}, we apply our framework to the
shot data of the 2015/16 season of the five big European
leagues to
(i) determine which players  significantly overperformed in terms of rGAX, (ii) empirically validate
the robustness of rGAX as opposed to GAX, and (iii) determine which goalkeepers
significantly overperformed in terms of shot-stopping.
We close our paper with a discussion and practical considerations in
Section~\ref{sec:disc}. The code to reproduce all analyses in this manuscript is
available at \url{https://github.com/Rob2208/rGAX_and_beyond}.

\section{Data and problem formulation}
\label{sec:problem-and-data}

A widely regarded strategy for gaining insight into shooting ability through
statistical methods is to start with an xG model. xG models assign a probability
of success to each shot, taking into account features describing the situational
context of each shot. Thereby, these models can be seen as a more nuanced
representation of a shot and allow for the quantification of the quality of a
scoring opportunity. In the following, we describe the problem of identifying
shooting skills, starting from the data structure and then following with a
common strategy for developing xG models and explain how they can be used to
evaluate a player's shooting ability.

\subsection{Data}
\label{sec:data}

\begin{figure}[t!]
    \centering
    \includegraphics[width=0.8\textwidth]{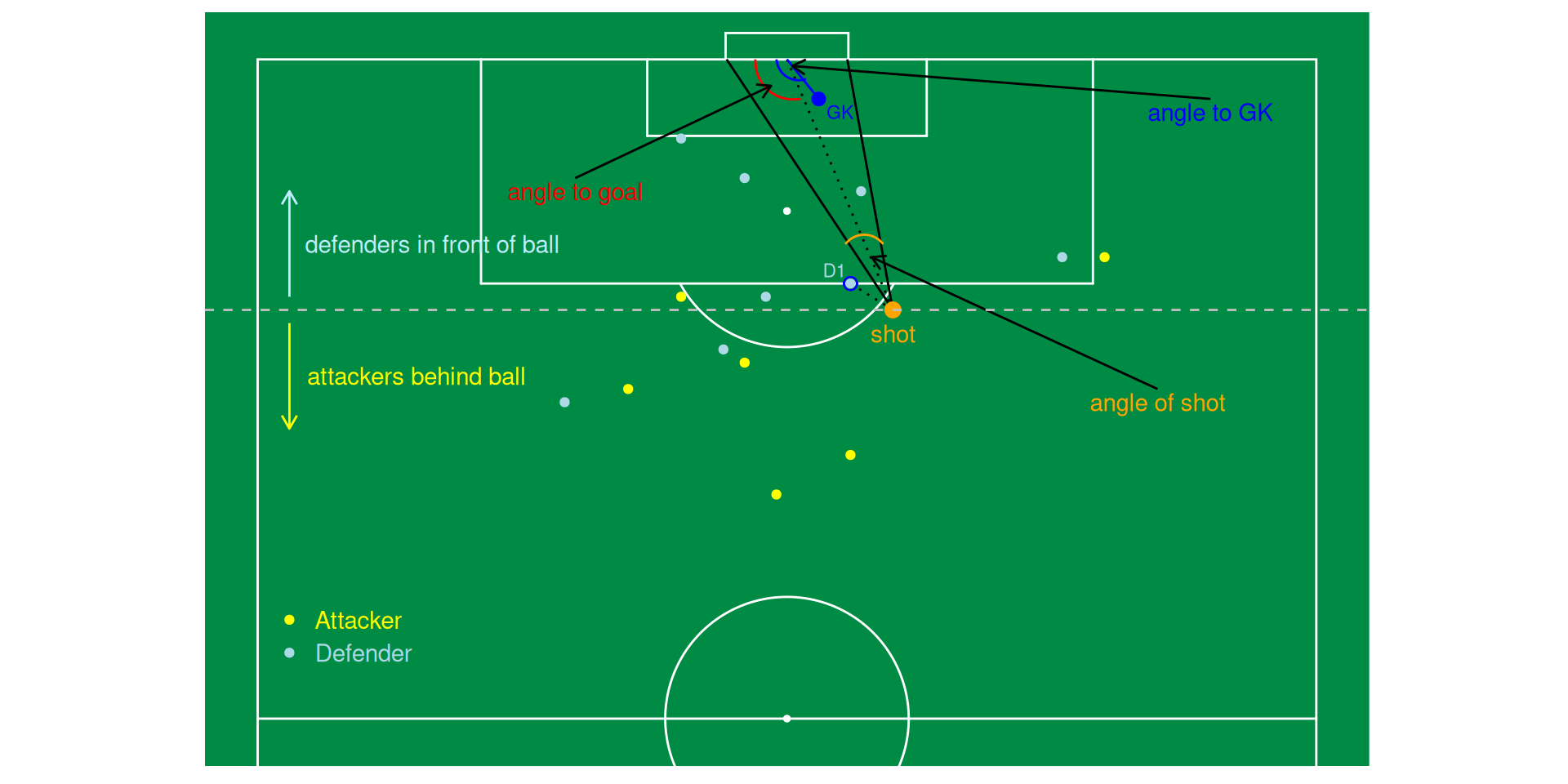}
    \caption{
    An instance of the data and the most important features used to compute xG
    models, GAX and rGAX. 
    }\label{fig:data}
\end{figure}

To obtain an xG model, we use event stream
data to obtain information on all shots from the 2015/16 season of the 2015/16
season of the Big Five European leagues (Bundesliga, La Liga, Ligue 1, Premier
League, and Serie A) provided by \href{https://statsbomb.com/}{Hudl-Statsbomb}
and obtained via the \pkg{StatsbombR} \proglang{R}~package
\citep{pkg:StatsBombR}. In particular, the data contain $N = 45197$ shots of
which 4308 resulted in a goal. Furthermore, the data contain information on the
shot location, the location of all other players that are visible by the camera
capturing the data, and a number of manually annotated information regarding the
shot, such as the shot type, the body part used for the shot, and the shot
technique. In concordance with previous work on xG models \citep[see,
e.g.,][among others]{RD20,AB21,HK23}, we extract a set of relevant shot-specific
features from the data, such as distance to the goal center, goalangle,
distances to defenders and the goalkeeper, and more. Figure~\ref{fig:data}
provides a snapshot of the most relevant features derived from the location of
the shot and the positions of the players. A full table of features for the xG
models can be found in Table~\ref{tab:features} in Appendix~\ref{app:data}.
Additionally, for each shot we observe whether the shot ended in a goal or not
and the name of the shooter.

%
The final dataset hence is comprised of $N$ observations
of a binary outcome variable of interest $Y \in \{0, 1\}$, representing
whether a shot ended in a goal (1) or not (0),
shot-specific features $Z$, and a categorical factor variable $F$ with
$P$ levels, where $P$ is the number of players in the dataset, identifying the
shooter associated with each shot. Using dummy encoding, for each player $p \in
\{1,\dots,P\}$, we can
create a dummy vector $X^p$ representing whether $p$ was the shooter (1) or not
(0).

\subsection{Evaluating shooting skill}
\label{sec:GAX_BG}

Given i.i.d.\ data $\{(Y_j, F_j, Z_j)\}_{j=1}^N$, an xG model
can be obtained by estimating the conditional expectation of $Y$ given
$Z$, $h(Z) \coloneqq \Ex[Y \given Z] = \Prob(Y = 1 \given  Z)$.
Typically, the literature does not directly account for player information in
the xG model, i.e.,~commonly used xG models do not condition on $F$. The
player information is then only used after fitting the xG model to evaluate
players (see details below). A traditional approach to estimate this conditional
expectation when $Y$ is binary, is to use a parametric logistic regression
model.
The logistic regression model assumes a linear relationship
between the log-odds of $h(Z)$ and the features $Z$: 
\begin{equation}\label{eq:xG_lrm}
\log\left(\frac{h(Z)}{1-h(Z)}\right) = Z^{\top}\gamma. 
\end{equation} 
This
modeling approach was chosen by \cite{PR97} for one of the earliest versions of
an xG model. As features describing the contextual characteristic of a shot,
\cite{PR97} used shot location, goalangle (defined as the angle between the
shot and the two goalposts), and an indicator for the body part used (foot or
head). More recently, xG models are trained on a broader set of features (such
as the features described above) and using machine learning algorithms such as
boosted tree ensembles \citep{AB21,HK23}, or neural networks
\citep{Corsaro2025xG}.

After estimating $h(Z)$ using a suitable model, such as, e.g., the logistic
regression model in \eqref{eq:xG_lrm}, a
 common strategy to evaluate the shooting ability of
player $p$ is then to compare the actual outcome of each shot of player $p$ to
the expected outcome given by the estimated model. That is, let $X^p_j$ denote
the binary indicator for whether player $p$ was the shooter of shot $j$, one is
interested in the empirical GAX of player $p$, defined as
\begin{equation}
\label{eq:GAX} 
\widehat{\operatorname{GAX}}_p \coloneqq \sum_{j=1}^{N} (Y_{j} - \hat
h(Z_{j}))X^p_j, 
\end{equation}
where $\hat h(Z)$ is an estimator of $h(Z)$ and corresponds to the xG value
for shot $j$. As mentioned in Section~\ref{sec:Intro}, while GAX are intuitive
and easy to explain, they recently have been subject to criticism
\citep{DR23xmskill,BSPC24}. In particular, \cite{DR24} state three main issues
with GAX: (1) the limited sample size for shots and goals leads to high
variances and unreliable estimates of shooting skill, (2) a bias arising from
including all shots (instead of only fractions of shots such as footers or
headers) obscures finishing ability, and (3) a bias arises from top teams and
top players taking more shots than weak teams or players. This paper aims to
address these issues by deriving a semiparametric approach for modeling shooting
skills, which (i) arises naturally by approaching the shooting skill problem
from a statistical angle (see Section \ref{subsec:lrm}), (ii) allows for a
deeper understanding and additional interpretability of GAX (see Section
\ref{sec:PLLM}), and (iii) is generalizable to a type of player evaluation
metrics commonly used in many sport domains (see Table~\ref{tab:applications},
and Appendix~\ref{app:use_cases}). Our proposed rGAX metric particularly
addresses the issues (2) and (3) from \citet{DR24} by additionally modeling the
propensity of a player for taking a shot given the circumstances of the shot
described by $Z$. Thereby, we implicitly account for the fact that top players
shoot more often, or are more likely to use a specific type of shot.
In particular, in Section~\ref{subsec:stability}, we find that rGAX are less
affected when estimating an xG model from data containing an overrepresentation
of a certain set of players. This supports the fact that our method is more
robust to
player biases in the data as opposed to traditional GAX. 
The limited sample size problem (1) is not easily
addressable, even when using rGAX. However, using rGAX allows for valid
uncertainty quantification of a player's shooting skill in form of confidence
intervals (see Section~\ref{subsec:GAX_vs_rGAX}).

\section{Methodological framework}

\subsection{A parametric approach to modeling shooting skills}
\label{subsec:lrm}

We motivate our semiparametric approach from a parametric modeling
perspective and present the generalization in Section~\ref{sec:PLLM}.
We work under a similar setup as before, where $Y$ is the binary outcome of a 
shot and $Z$ are shot-specific features. Since we are interested in evaluating
the shooting ability of player $p$, we
additionally add a binary variable $X^p$
to the data, indicating whether the player of interest was the shooter
of the shot (1) or not (0). As mentioned previously, a traditional parametric
approach for modeling binary outcome data is a logistic regression model. 
That is, $Y \given  X^p,Z \sim
\operatorname{Ber}\left({h_p}(X^p,Z)\right)$ follows a Bernoulli
distribution with
${h_p}(X^p,Z) \coloneqq \Ex[Y \given X^p,Z] =~ \mathbb{P}(Y=1 \given  X^p,Z)$, and
\begin{equation}\label{eq:lrm}
\log\left(\frac{{h_p}(X^p,Z)}{1-{h_p}(X^p,Z)}\right) = X^p\beta + Z^{\top}\gamma.
\end{equation}
The parameter of interest in this setup is $\beta$, which can be interpreted as
a player's effect on the log odds (or probability) of scoring. Statistical
inference on $\beta$ in this model has been well understood for many years,
and one popular approach is to use a score test \citep{Rao48}. Given
i.i.d.~data $(Y_j,X^p_j,Z_j)_{j = 1}^N$ from the above logistic regression model,
the test targets the score of $\beta$, defined as 
\begin{equation}\label{eq:score}
\sum_{j = 1}^N\frac{\partial\ell(\beta,\gamma \given  Y_j,X^p_j,Z_j)}{\partial
\beta},
\end{equation}
where $\ell(\beta,\gamma \given  Y,X^p,Z)$ 
denotes the log-likelihood function of the
logistic regression model
\begin{equation}\label{eq:ll_lrm}
\ell(\beta,\gamma \given  Y,X^p,Z) \coloneqq\sum_{j = 1}^N Y_j
\log\left({h_p}(X^p_j,Z_j)\right)+(1-Y_j)\log\left(1-{h_p}(X^p_j,Z_j)\right).
\end{equation}
Under the null hypothesis of interest $H_0: \beta = 0$, the score can be 
computed as
\begin{equation}\label{eq:score_lrm}
\sum_{j=1}^{N} (Y_j - \hat h(Z_j))X^p_j,
\end{equation}
where $\hat h(Z_j) = \text{expit} (Z_j^{\top}\hat \gamma)$ 
is an estimator of $h(Z) \coloneqq \Ex[Y \given Z]$
(which, under $H_0$, no longer depends on $X^p$), and $\hat \gamma$
denotes the maximum likelihood estimate of the (vector-valued) parameter
$\gamma$ under $H_0$. Since $X^p_j$ is a binary variable that is only one when the
player of interest $p$ was the shooter of shot $j$, GAX is exactly the score from a
logistic regression model under $H_0$, when using a logistic regression model to
fit the xG model. This connection allows for a deeper understanding of GAX and
additionally provides a new interpretation. On the one hand, we see that GAX is
intimately related to the score of a player's effect parameter, thereby allowing
for valid uncertainty quantification and significance testing in model
\eqref{eq:lrm}. On the other hand, instead of trying to interpret GAX, we can
equivalently analyze the coefficient $\beta$ from \eqref{eq:lrm}, i.e.~the
effect of a player on the log odds of scoring a goal from a shot, while
accounting for the circumstances $Z$ of the shot.

Although the above connection reveals interesting insights into GAX, several
problems remain. First, the linear model assumptions underlying the logistic
regression model are unrealistic and do not capture the complexity of shots.
This problem is backed up by the literature on xG models, which suggests that
flexible non-linear machine learning models outperform the classical logistic
regression model \citep{RD20,AB21}. 
Furthermore, traditional xG models consider only shot-specific
variables, leading to potential biases by ignoring contextual factors such as
team strengths, goalkeeper strengths, and other potential player-specific
effects, as criticized by \citet{DR24}. Accounting for these additional factors
in the logistic regression model increases the problem's dimensionality
drastically, potentially invalidating the inference on player effects. Finally,
using the modern approach
of computing GAX by learning the xG model via machine learning methods,
i.e.~considering a score of the form
\begin{equation}\label{eq:GAX_ml}
\sum_{j=1}^{N} (Y_j - \hat h(Z_j))X^p_j,
\end{equation}
where $\hat h$ is estimated via an arbitrary machine learning algorithm,
yet no longer allows for valid parametric inference.

\subsection{rGAX: A semiparametric framework for player evaluation}
\label{sec:PLLM}

A natural extension of the model in~\eqref{eq:lrm} from Section~\ref{subsec:lrm}
is the partially linear logistic model (PLLM). In this semiparametric model, the
binary outcome variable $Y \given X^p, Z \sim
\operatorname{Ber}({h_p}(X^p,Z))$
follows a Bernoulli distribution with ${h_p}(X^p,Z) =
\mathbb{P}(Y=1 \given  X^p,Z) = \Ex[Y \given
X^p,Z]$, and 
\begin{align}
\label{eq:pllm_lo}
\begin{aligned}
\log\left(\frac{{h_p}(X^p,Z)}{1-{h_p}(X^p,Z)}\right) =
X^p\beta + g(Z),
\end{aligned}
\end{align}
with some arbitrary measurable function $g$. The parameter of interest
$\beta$ linearly influences the log odds for a positive outcome (i.e.~a goal
from a shot) and hence, 
the interpretation of $\beta$ is exactly the same as in the
parametric logistic regression in Section~\ref{subsec:lrm}.
Next, we discuss how to achieve valid statistical inference for $H_0 :
\beta = 0$.

For inference on $\beta$, we will rely on the recently developed
Generalised Covariance Measure (GCM) due to \citet{SP20}.
The GCM test targets
the expected conditional covariance between $Y$ and $X^p$ given $Z$:
\begin{equation}\label{eq:GCM}
\operatorname{GCM}_p \coloneqq
\mathbb{E}[\operatorname{Cov}(Y,X^p \given  Z)] =\mathbb{E}[(Y -
\mathbb{E}[Y \given Z])(X^p - \mathbb{E}[X^p \given Z])].
\end{equation}
The basis for the test is that a necessary condition for conditional
independence of $Y$ and $X^p$ given $Z$, denoted by $Y
\perp\!\!\!\perp X^p \given  Z$,
is that $\mathbb{E}[\operatorname{Cov}(Y,X^p \given  Z)] = 0$.
In practice, to use the GCM test, a sample version of the GCM needs to be
estimated. From the second representation in \eqref{eq:GCM}, it can be
seen that this is achieved by learning two regression functions $h(Z) =
\mathbb{E}[Y \given Z]$ and $f_p(Z) \coloneqq \mathbb{E}[X^p \given
Z]$ (using the same features $Z$ for both). 
In particular, \citet{SP20} show that under mild rate conditions akin to
conditions in debiased machine learning \citep{chern2017double}, which can
typically be achieved by modern machine learning algorithms, the sample version
of the GCM
\begin{equation}\label{eq:GCM_sample}
\widehat{\operatorname{GCM}}_p \coloneqq \frac{1}{N}\sum_{j = 1}^N (Y_j-\hat
h(Z_j))(X^p_j
- \hat f_p(Z_j))
\end{equation}
is asymptotically normal with variance shrinking at rate $1/N$.
The variance of this normal distribution can be consistently estimated by the
empirical variance of $\widehat{\operatorname{GCM}}_p$. 
More precisely, for the GCM test to be valid, the product of the average squared
deviations of the estimated regressions functions from their ground truths
needs to vanish at a rate of $1/N$,
i.e.,
\begin{equation}\label{eq:GCM_cond}
\frac{1}{N} \sum_{j = 1}^N (h(Z_j)-\hat{h}(Z_j))^2 
\cdot
\frac{1}{N} \sum_{j = 1}^N (f_p(Z_j)-{\hat{f}_p}(Z_j))^2
= o_P(N^{-1}),
\end{equation}
where $a_N = o_P(b_N)$ when $a_N/b_N \to 0$ in probability. 
A similar set of conditions is employed for the estimation of causal parameters
via double machine learning \citep{chern2017double} and assumption lean
inference on generalized linear model parameters \citep{VD22ALR}. The condition
on the product error for the regression implies that the test is valid even if
both regression functions are learned at a nonparametric rate. In this context,
``doubly robust'' refers to the case that if one regression ($\hat h$ or 
$\hat{f}_p$) is estimated at a sufficiently fast rate (e.g. $ \frac{1}{N} \sum_{j = 1}^N
(h(Z_j)-\hat{h}(Z_j))^2 = o_P(N^{-1})$), the other regression can be much less
accurate, and the product condition can still hold. In simpler terms, if at
least one of the learned regression functions $\hat h$ and $\hat f_p$ approximate
the true functions $h$ and $f_p$ well enough, valid inference and uncertainty
quantification for the GCM is possible, allowing for tests of conditional
independence between $Y$ and $X^p$ given $Z$. Although the GCM does not require
sample splitting for the theoretical guarantees to hold, careful tuning of the
regressions for obtaining $\hat h$ and $\hat f_p$ is advised
\citep{SP20,kook24comets}.

The following results connect the GCM test to a test on the parameter
$\beta$ in the PLLM in \eqref{eq:pllm_lo}.
\begin{proposition}
\label{thm:prp1}
Let $(Y, X^p, Z)$ take values in $\{0, 1\} \times \{0, 1\} \times
\RR^{d_Z}$ with distribution $P$, such that there exist a
$P_Z$-almost surely finite
function $g : \RR^{d_Z} \to \RR$, and $\beta \in \RR$ such that 
the partially linear logistic model in~\eqref{eq:pllm_lo} holds with 
$0 < P(X^p = 1 \given Z) < 1$ $P_Z$-almost surely. Then, the
following two statements hold:
\begin{itemize}
    \item[(i)] $\beta = 0$ if and only if $\mathbb{E}[\operatorname{Cov}(Y,X^p
    \given  Z)] = 0$,
    \item[(ii)] $\operatorname{sign}(\beta) =
    \operatorname{sign}(\mathbb{E}[\operatorname{Cov}(Y,X^p \given  Z)])$.
\end{itemize}
\end{proposition}
The proof of Proposition~\ref{thm:prp1} can be found in Appendix~\ref{app:proofs}.
Proposition~\ref{thm:prp1} entails that we can use the GCM
test for testing the hypothesis $H_0 : \beta = 0$ in the PLLM. This is very
convenient and allows for model agnostic testing in our setup, i.e.~testing
without imposing any (parametric) model constraints. The only
requirement for the GCM test is that the rates of the machine learning models
used for the estimation of $h$ and $f_p$ are fast enough. This can be achieved
by a properly tuned machine learning algorithm tailored to the problem at hand. 
Additionally, Proposition~\ref{thm:prp1} $(ii)$ entails that the GCM allows for
directional testing, e.g.,~testing hypotheses of the form $H_A : \beta > 0$. In
terms of interpretation, this means that the outcome of the GCM test is related
to a ``strength'' estimate of a player in the PLLM, allowing us to infer players
having a significant positive impact on the probability of scoring. 

Finally, we can also connect the GCM test and therefore the parameter $\beta$ of
the PLLM to GAX. Recall that 
traditional empirical GAX for  player $p$ can be written as
\begin{equation*}
\widehat{\operatorname{GAX}}_p = \sum_{j=1}^{N} (Y_j - \hat h(Z_j))X^p_j,
\end{equation*}
where $\hat h$ is an arbitrary estimate for $h(Z) = \mathbb{E}[Y \given Z]$ and
corresponds to an xG model. If we use a logistic regression model as xG model,
we obtain a GAX value that allows for valid uncertainty quantification and can be
related to a strength estimate in a parametric model. However, as has been
pointed out repeatedly, flexible machine learning models are more suitable for
capturing the complex relationship between outcome of a shot and shot-specific
features. Using a machine learning model, valid inference is, however, no longer
guaranteed.
To address this issue, we propose to use empirical residualized GAX (rGAX)
\begin{equation}
\label{eq:rGAX}
\widehat{\operatorname{rGAX}}_p \coloneqq \sum_{j=1}^{N} (Y_j-\hat
h(Z_j))(X^p_j - \hat f_p(Z_j)),
\end{equation}
a scaled version of the sample GCM, for the dependence between $Y$ and $X^p$ given $Z$.

Both $\hat{h}$ and $\hat{f}_p$ are estimated using the same feature set $Z$ —
that is, the shot-specific features described in Section~\ref{sec:data}, with
the key distinction that $Y$ (goal or no goal) and $X^p$ (whether player $p$
took the shot) serve as the respective response variables.

rGAX is defined  on the same scale as GAX (so rGAX and GAX are directly
comparable) and has several advantages over classical GAX. First, in comparison
to a
machine learning based GAX, rGAX allows for valid inference combined with
intuitive interpretation as a strength estimate on the log-odds of scoring a
goal from a shot via the $\beta$ coefficient in a PLLM.
Second, 
rGAX requires estimating the additional regression function $f_p(Z)$.
Estimating this function is particularly interesting from a statistical viewpoint,
because it enables valid inference for rGAX even when using machine learning
models to estimate $h(Z)$ and $f_p(Z)$. However,
the additional
regression of $X^p$ on $Z$ also has a domain-specific interpretation:
For each shot
that we have in our data, we model the propensity of a specific player taking
the shot, given the circumstances $Z$ of the shot. That is, instead of
considering shots taken as in GAX, we consider whether a player would be likely
to take a shot. 
This aligns with the widely recognized practical understanding in soccer
that players exhibit individual-specific shooting tendencies and may perform
particularly well in situations that match their preferred techniques, even when
such situations are comparatively difficult on average.\footnote{A well-known
example is Arjen Robben, whose tendency to cut inside from the right flank and
shoot with his left foot toward the far corner became a characteristic and
highly effective attacking pattern. Although this type of shot is generally
considered challenging, Robben executed it with exceptional effectiveness.}
Consequently, $f_p(Z)$ does not necessarily capture the difficulty of
a shot.
Instead, it reflects the extent to which a given shot situation matches a
player's characteristic strengths. Shot difficulty is primarily captured by
$h(Z)$. 

Third, for the regression of
$Y$ on $Z$, any given available xG model can be used. This has two
advantages: On the one hand, to compute rGAX, it is not necessary to fit a new
xG model, but one can rely on already trained models. On the other hand, using
the same xG model for GAX and rGAX allows for a fair comparison of the two
metrics. Lastly, as rGAX arise from the test statistic of the GCM test,
an inherently non-parametric conditional independence test, it is a meaningful
statistic even in the case that the modeling assumptions of~\eqref{eq:pllm_lo}
are not met. That is, rGAX and the corresponding test results ($p$-values and
confidence intervals) can be meaningfully used to measure a player's shooting
quality in a much more general setup than the PLLM model. The advantage
of~\eqref{eq:pllm_lo} is, however, twofold. It directly motivates the usage of
rGAX as a straightforward extension of GAX obtained via a parametric framework.
And secondly, it allows for the previously discussed appealing and
straightforward interpretation of $\beta$. 
A code example for applying our
framework can be found in
Appendix~\ref{app:code}.

\subsection{Evaluating goalkeepers}
\label{subsec:GSAX}

A related problem to evaluating shooting skills is the evaluation of goalkeepers. 
A popular approach for measuring goalkeeper skill is to use goals saved above
expectation (GSAX). In fact, GSAX is similar to GAX, however
only for shots on target, as goalkeepers should only be evaluated on these.
Additionally, the target of goalkeepers is not to score but to prevent scoring from
a shot. Thus, an outstanding goalkeeper should have a significant negative
impact on the probability of scoring from a shot.
In more detail, to evaluate GSAX, a post-shot expected goals (psxG)
model is learned, taking into account only shots on target as well as
information after the actual shot was taken, such as the shot's trajectory and end
location. That is, for a given goalkeeper $p$ (encoded through the indicator $X^p_j$),
empirical GSAX are defined as
\begin{equation}\label{eq:GSAX}
\widehat{\operatorname{GSAX}}_p \coloneqq -\sum_{j=1}^{N} (Y_j - \hat h(Z_j))X^p_j,
\end{equation}
where $Y_j$ represents the actual outcome of shot $j$ on target, $\hat h(Z)$
is an estimator for $h(Z) = \mathbb{E}[Y\given Z]$ and corresponds to the psxG value
for shot $j$, and $Z_j$ contains information on the shot trajectory. Analogous to
GAX, we propose to use rGSAX instead of GSAX, the empirical version defined as
\begin{equation}
\label{eq:rGSAX}
\widehat{\operatorname{rGSAX}}_p \coloneqq -\sum_{j=1}^{N} (Y_j-\hat
h(Z_j))(X^p_j - \hat f_p(Z_j)).
\end{equation}
Notably, the equation for rGSAX is almost the same as \eqref{eq:rGAX} 
for rGAX with two main differences: First, we only consider 
shots on targets and thus $Z$
contains more (or different) information than before. In this case, $\hat h$ is
the learned psxG model and $\hat f_p$ is a model that accounts for the propensity
of a goalkeeper to face a shot given the circumstances $Z$. This additional
regression can again be interpreted from a domain-specific viewpoint: Instead of
considering actual shots faced by the goalkeeper, we consider whether a
goalkeeper would be likely to face a shot given the circumstances $Z$ of the
shot. This accounts for the quality of goalkeepers, as good goalkeepers may be
able to anticipate shots better than average goalkeepers. 
Further, the resulting rGSAX can again be related to a
valid significance test, and the result can be interpreted as the impact a
goalkeeper has on the log-odds (and therefore probability) of scoring a goal
in a PLLM. In contrast to the shooting ability of a player, one would, in this
case, be interested in whether a goalkeeper has a significant negative effect on
the log-odds of scoring a goal from a shot. For the sake of consistency in the
results, we hence add a negative sign in \eqref{eq:rGSAX}. Therefore, similar to
the case of GAX, a positive GSAX value represents strong goalkeepers. 

\section{Results}
\label{sec:res}

\subsection{Additional detail on the data}
\label{sec:add_data}

In this section, we present the results of the proposed framework and compare
rGAX to GAX. We use the data described in Section~\ref{sec:data} and
Figure~\ref{fig:data}. Additionally, in order to account for team quality, we
derive a defensive strength parameter for the opposing team on each shot. To do
so, we use a bivariate Poisson model on match outcome data \citep{KN03}. Details
on this procedure can be found in Appendix~\ref{subsec:ts}.  We evaluate GAX and
rGAX for all players who shot at least 20 times during the observed season and
scored at least one goal. This amounts to 728 players. 

To analyze goalkeepers via GSAX and rGSAX, we use the same data as described
above. However, for goalkeeper analyses, it is only sensible to consider shots on
target. For all shots, the data contain information on the end location on the
$y$ and $z$ plane. Since the dimensions of a goal are fixed, we can derive shots
on target by using the shot end location information. In total, we end up with
13269 shots on target to analyze. We fit the psxG model using the same features
as for the xG model, with the addition of two features indicating the $y$ and
$z$ difference to the center of the goal. We again account for team quality by
deriving an attacking strength parameter for the team shooting the ball via a
bivariate Poisson model. Similar to before, we only consider goalkeepers who
were on the field for at least 20 shots against them, resulting in 147
goalkeepers to analyze.

\subsection{Computational details} \label{sec:comp}

For the computation of the xG models, the GCM test, and to obtain estimates of
GAX, rGAX, GSAX, and rGSAX, we use the \proglang{R} language for statistical
computing \citep{RLang25} and the \proglang{R}~package \pkg{comets}
\citep{kook24comets}. The \pkg{comets} package allows for fitting the
regressions of $Y$ and $X^p$, $p = 1, \ldots, P$, on $Z$ via
a broad range of flexible machine learning models and simultaneously tuning
them. Furthermore, in \pkg{comets} it is possible to use pre-trained models for
the regressions, allowing us to use an already fitted xG model for the
regression of $Y$ on $Z$. Hence, we have a fair comparison of GAX and rGAX by
using the same xG model for both. Similarly, for GSAX and rGSAX, we use the same
pre-trained psxG model.

Following existing literature  \citep{RD20,AB21,HK23}, we use boosted regression
trees to fit an xG model. The \pkg{comets} package can conveniently obtain these
via the \proglang{R}~package \pkg{xgboost} \citep{xgboost25}. We use a similar
methodology for the psxG models. Furthermore, for both models, we perform
extensive cross validation on a grid of values for tuning the hyperparameters.
For the regression of $X^p$ on $Z$, we use an out-of-bag tuned random
forest using the \proglang{R}~package \pkg{ranger} \citep{Wright17ranger}. We
discuss several choices for the regression and potential problems arising from
under- and overfitting in Appendix~\ref{app:rGAX_add}. We give details on the
hyperparameter tuning in Appendix~\ref{subsec:hyperpar}.

\subsection[Evaluating shooting skill: GAX vs rGAX]{Evaluating shooting skill:
GAX \emph{vs}.\ rGAX}\label{subsec:GAX_vs_rGAX}

\begin{figure}[t!]
    \centering
    \includegraphics[width=0.95\textwidth]{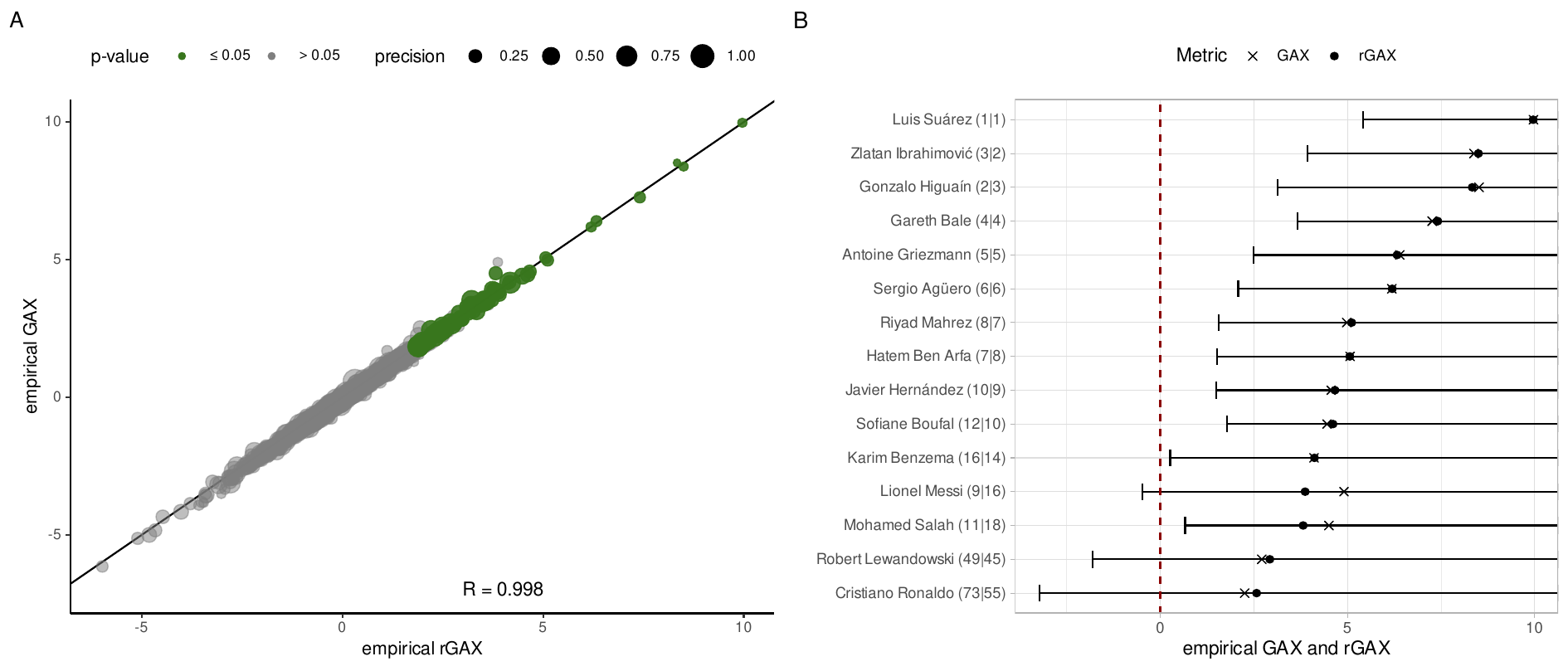}
    \caption{
    Goals and residualized goals above expectation plots for the 2015/16 season
    of the 5 big European soccer leagues. A: Scatterplot of empirical GAX and
    rGAX. The size of the dots represents the estimated precision of empirical
    rGAX, defined as the inverse of its estimated variance.
    The solid line indicates the identity. The correlation coefficient $R$
    is added to the plot. B: Player-wise empirical GAX and rGAX with one-sided
    95\% confidence intervals for rGAX (without multiplicity corrections)
    for the top 10 players in terms of empirical rGAX and 5 selected well-known
    players. The rank of the players in term of GAX (left) and rGAX
    (right) is shown in brackets.
    }
    \label{fig:GAX_vs_rGAX}
\end{figure}

We computed empirical GAX and rGAX for all 728 players in our data, and display
the results in Figure~\ref{fig:GAX_vs_rGAX}. Figure~\ref{fig:GAX_vs_rGAX}A shows
a scatterplot of empirical GAX and rGAX fitted as described in
Section~\ref{sec:comp}. The high correlation between empirical GAX and rGAX
indicates that both metrics measure shooting skills similarly. This is desirable
as rGAX can and should be interpreted in the same manner as the commonly used
GAX. Additionally, this suggests that a suitably designed xG model used to
compute (empirical) GAX may be able to adequately capture shooting skill. rGAX,
however, allow for more insights: A major advantage of rGAX is that they enable
statistical uncertainty quantification by computing confidence intervals and
$p$-values. In Figure~\ref{fig:GAX_vs_rGAX}B, we show the top ten players ranked
by empirical rGAX as well as five well-known players. The one-sided confidence
intervals for the 15 players are shown in that figure. Hence, we are able to
identify which players' rGAX are significantly greater than 0 (at the 5\%
significance level). This is not only desirable from a statistical perspective
but also opens up new possibilities for evaluating the stability of GAX. Instead
of the classical approach of comparing empirical GAX (or rGAX) from one season
to the next \citep{BSPC24}, it is possible to analyze $p$-values and confidence
intervals for rGAX over various seasons. We cannot directly test/address this in
this manuscript due to limited data availability. Furthermore, we can interpret
these values in terms of the PLLM of Section~\ref{sec:PLLM}.
Figure~\ref{fig:GAX_vs_rGAX}B shows that for all of the top ten players
according to (empirical) rGAX, the one-sided confidence intervals do not
contain the value 0. Relating rGAX to the PLLM via Proposition \ref{thm:prp1},
this suggests that these players have a statistically significant positive
effect on the probability of scoring a goal when shooting. The results can also
be interpreted from a domain-specific viewpoint. Most of the top ten players are
well-known strikers in the top leagues of Europe. A more surprising result shows
that Lionel Messi and Cristiano Ronaldo, the two players widely regarded as the
best players at that time, while still having comparably high empirical rGAX
values (ranking in the top 60 from 728 players), do not achieve a value of rGAX
which is significant at conventional levels. Also the wide confidence interval
for the rGAX of L.\ Messi stands out. One possible explanation is that in the
2015/16 La Liga season, L.\ Messi performed rather poorly in terms of
traditional shooting metrics, with him scoring the second lowest number of goals
compared to all other seasons from 2009 onwards. While the results are
interesting and allow for a much deeper insight into the analysis of soccer
player shooting skills than GAX, they have to be interpreted with care. On the
one hand, rGAX only allow for identifying shooting skills of soccer players, but
there may be other aspects determining outstanding soccer players. On the other
hand, there are a number of potential practical considerations which relate to
the underlying assumptions of using (r)GAX: The independence assumption
necessary for valid inference, small effective sample size due to the limited
number of goals, the need for multiplicity corrections, and the choice of
control variables for the regression models. In Section~\ref{subsec:pc}, we
discuss these practical considerations when interpreting the results in more
detail.    

\subsection{Robustness of GAX and rGAX}
\label{subsec:stability}
A major concern of \cite{DR24} for using GAX to measure shooting skills is the
inherent selection bias problem in soccer. \cite{DR24} point out that, due to the
nature of the game, we observe more shots from strong shooters (and strong teams)
as opposed to weak players (and weak teams). Hence, fitting an xG model on these
data to obtain GAX induces a bias. 
rGAX, however, are able to account for 
this bias due to estimating the propensity of a player taking a shot under the
given circumstances of the shot described by $Z$. 

To illustrate this point, we pick up on the example of \cite{DR24} and consider
a dataset with an overrepresentation of shots from Lionel Messi. In 
particular, in addition to the 2015/16 data of the top 5 European leagues,
\href{https://statsbomb.com/}{Hudl-Statsbomb} also openly provides a biography
of all shots of Lionel Messi during his time at FC Barcelona via the
\pkg{StatsBombR} package. This dataset 
contains 2499 shots of Lionel 
Messi, with the next frequent shooter (Luis Suárez, a long-term 
teammate of Messi) only having 607 shots. As Messi is widely considered a
top shooter with a unique shooting profile, fitting an xG model to these
data does not accurately capture the average player, and therefore GAX
values of players may be obscured. 

In our experiments, we augment the 2015/16 data with the Messi shot data 
and fit three xG models. One model using all the data containing the
overrepresented Messi shots. A second model using the 2015/16 data as 
in the previous sections. These shots are more equally balanced, but may still 
contain an overrepresentation of good players taking more shots. Finally, 
we fit a third xG model using only shots from players with at most 30 shots
observed (low-frequency model). Thereby, this model serves as a proxy for a 
model representing low-quality shooters. We use the three models to compute 
empirical GAX and rGAX for
all players with at least 70 shots in the augmented data,
resulting in a total of 136 players.   

\begin{figure}[t!]
    \centering
    \includegraphics[width=0.8\textwidth]{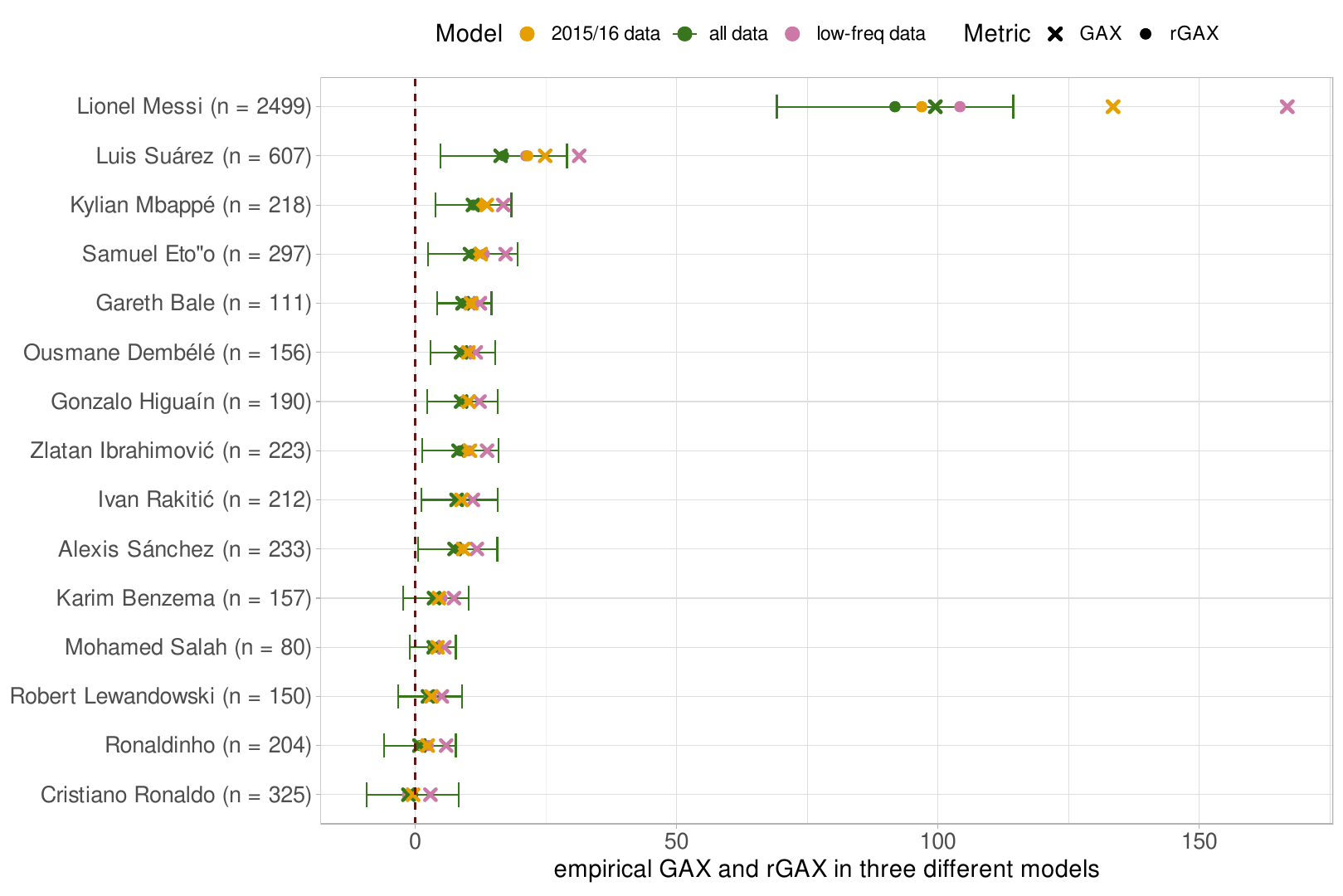}
    \caption{
    Player-wise empirical GAX and rGAX computed from three different xG models
    for the top 10 players with respect to empirical rGAX  and 5 selected
    well-known players. The 95\% confidence intervals for rGAX (without
    multiplicity corrections) from the model using all data are shown. 
    }
    \label{fig:rGAX_stability_players}
\end{figure}

Figure~\ref{fig:rGAX_stability_players} displays the resulting 
empirical GAX and rGAX
values for the top 10 players, as well as 5 selected well-known players. 
The results show that Messi's empirical GAX and rGAX values are highly positive,
outperforming every other player by far, but also having taken more than four
times the amount of shots than the next player. Additionally, 
we observe that GAX are much more 
drastically affected by the particular
xG model used to compute them. In particular, the 
empirical GAX value for all
players are shifted to the right, i.e., overestimated, when using the models
trained on less data, with an especially pronounced effect when computing 
empirical GAX from the low-frequency model. rGAX,
on the other hand, are affected less
drastically by the choice of model and data.

\begin{figure}[t!]
    \centering
    \includegraphics[width=0.95\textwidth]{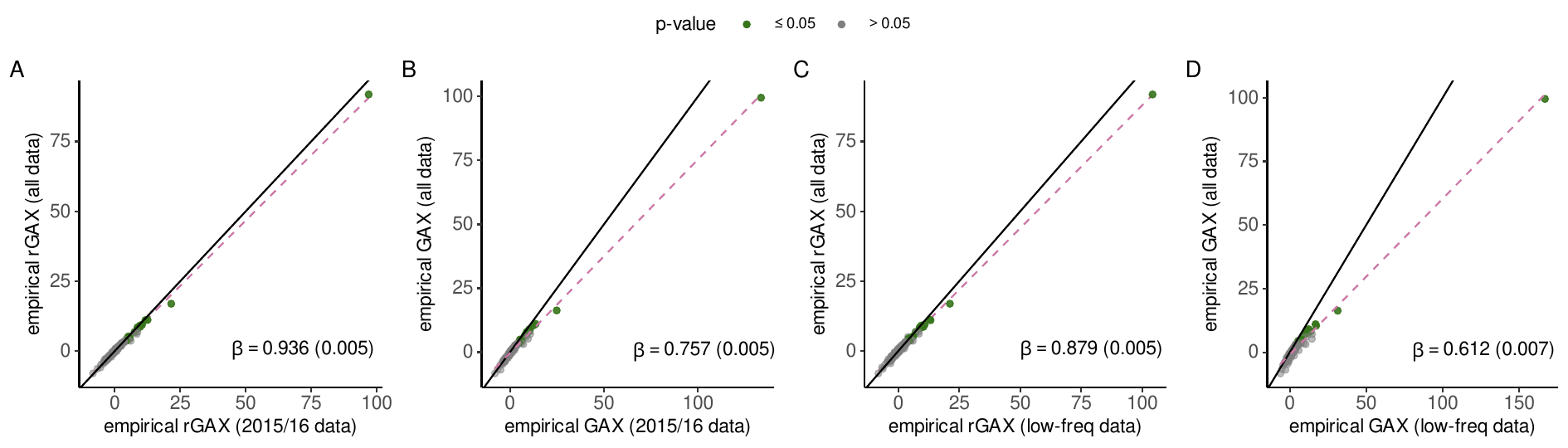}
    \caption{
    Scatterplots of empirical GAX and rGAX values computed from different models. 
    A: Scatterplot for empirical rGAX from an xG model computed with all data available
    against empirical rGAX from a model using only 2015/16 data. 
    B: Scatterplot for empirical GAX from an xG model computed with all data
    available against empirical GAX from a model using only 2015/16 data 
    C: Scatterplot for empirical rGAX from an xG model computed
    with all data available against empirical rGAX from a model using 
    only low-frequency shooter data. 
    D: Scatterplot for empirical GAX from an xG model computed with all data
    available against empirical GAX from a model using only low-frequency 
    shooter data.
    Solid black lines correspond to the identity, and dashed red lines
    correspond to linear regression fits on the data.
    The coefficient $\beta$ of 
    a linear model with intercept estimated from the data (and its standard
    deviation) is shown in each plot.
    }
    \label{fig:rGAX_stability}
\end{figure}

Figure~\ref{fig:rGAX_stability} reinforces the findings from Figure
\ref{fig:rGAX_stability_players}. The figure shows scatterplots of empirical
rGAX values from different models and empirical GAX values from different
models. Figures~\ref{fig:rGAX_stability}A and~\ref{fig:rGAX_stability}C display
the empirical rGAX values computed from the xG model using 2015/16 data only and
the low-frequency shooter data, respectively, against the empirical rGAX values
from the xG model using all data. Figures~\ref{fig:rGAX_stability}B and
\ref{fig:rGAX_stability}D similarly show the corresponding empirical GAX values.
In all four plots, the solid black line indicates the identity, whereas the
dashed red line denotes a regression line for the data. The smaller deviance
between the black and red lines in Figures~\ref{fig:rGAX_stability}A and
\ref{fig:rGAX_stability}C as opposed to~\ref{fig:rGAX_stability}B and
\ref{fig:rGAX_stability}D demonstrate the higher variability in (empirical) GAX
as opposed to (empirical) rGAX. Hence, GAX are more dependent on the choice of
data used to compute the xG model, whereas rGAX are more robust due to
additionally modeling a player's propensity to take a certain shot. 

\subsection[Evaluating goalkeepers: GSAX vs. rGSAX]{Evaluating goalkeepers: GSAX
\emph{vs}.\ rGSAX}\label{subsec:GSAX_vs_rGSAX}

\begin{figure}[t!]
    \centering
    \includegraphics[width=0.95\textwidth]{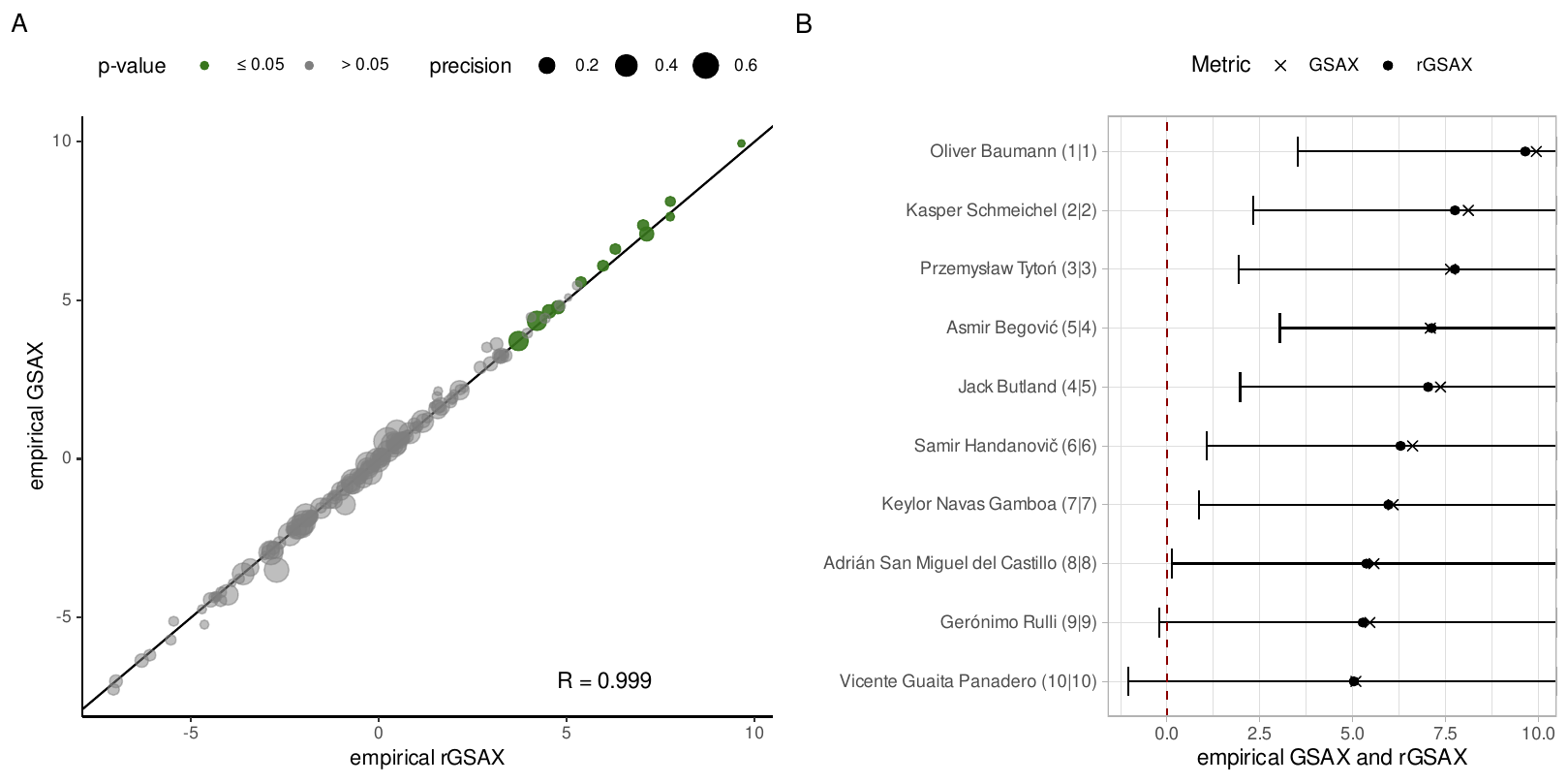}
    \caption{
    Goals saved and residualized goals saved above expectation plots for the
    2015/16 season of the 5 big European soccer leagues. A: Scatterplot of
    empirical GSAX and rGSAX. The size of the dots represents the
    estimated precision of empirical rGSAX. The solid line indicates the identity. The
    correlation coefficient $R$ is added to the plot. B: Player-wise empirical
    GSAX and rGSAX with one-sided 95\% confidence intervals for rGSAX
    (without multiplicity corrections) for the top 10 players in terms
    of empirical rGSAX. The rank of the players in terms of GSAX (left)
    and rGSAX (right) is shown in brackets.
    }
    \label{fig:GSAX_vs_rGSAX}
\end{figure}

We evaluate the shot-stopping skills of goalkeepers similar to shooting skills
in Figure~\ref{fig:GSAX_vs_rGSAX}. Figure~\ref{fig:GSAX_vs_rGSAX}A plots
empirical GSAX and rGSAX against each other for the 147 goalkeepers in our data.
Similar to before, we observe near-perfect correlation between empirical GSAX
and rGSAX. As opposed to GSAX, rGSAX again allows for uncertainty quantification
and interpretation in our semiparametric framework. Figure
\ref{fig:GSAX_vs_rGSAX}B displays the top 10 goalkeepers ranked by empirical
rGSAX together with the one-sided 95\% confidence interval for rGSAX. The figure
suggests that the top 8 goalkeepers have a statistically significant negative
effect on the likelihood of scoring a goal. That is, the results indicate that
there is statistical evidence that these players have a negative effect on the
outcome of scoring a goal from a shot when being the goalkeeper against a shot.

\section{Discussion and practical considerations}
\label{sec:disc}

In this work, we present a generalization for a common type of player evaluation
metrics in sports that can be expressed as score statistics from a parametric
model. Our main focus is on GAX, i.e., the difference between actual and
predicted goals from shots of a player, a metric commonly used in soccer to
evaluate shooting skills and a prime example of our framework. We show that GAX
naturally arises as a score statistic of a player strength estimate in a
parametric model. To allow for more flexibility, we provide an extension of GAX
to a player strength estimate in a semiparametric model. For player evaluation,
we propose rGAX, a metric arising from the player strength estimate in the
semiparametric model that is directly comparable to GAX. We apply our framework
to the 2015/16 season of the top five European leagues to determine the best
shooters.

Our results show that GAX and rGAX, in essence, measure shooting skills
similarly, as evidenced by near-perfect correlation between the two metrics (see
Figure~\ref{fig:GAX_vs_rGAX}). However, rGAX have two main advantages over GAX:
They allow (i) for valid frequentist uncertainty quantification, i.e., for the
computation of valid confidence intervals and $p$-values, and (ii) for the
interpretation of rGAX as a player strength estimate in a semiparametric model,
i.e., rGAX are related to the effect a player has on the likelihood of scoring a
goal from a shot while accounting for the shot-specific circumstances $Z$. 

Additionally, rGAX address recent criticism of GAX as a measure of shooting
skill (see Section~\ref{sec:GAX_BG}). In particular, \cite{DR24} criticize GAX
for not adequately measuring skills due to (1) limited sample size resulting in
high variances, (2) biases arising from including all shots as opposed to only
specific categories (e.g., footed shots), and (3) biases arising from
interdependencies in the data (selection bias). The issue of limited sample size
is innate to soccer, as shots and goals scored are only rarely observed events.
Hence, it is difficult to directly address or circumvent the issue. However, the
ability of rGAX to provide valid frequentist inference allows to quantify the
variability in the estimates. We view (2) as less of an issue of GAX per se, but
a problem of defining the correct estimand of interest. To pick up on an example
of \cite{DR24}, we agree that taking into account footed and headed shots may
make a difference for players particularly good at one category. rGAX are able
to address this issue by accounting for the shot type (e.g., footed vs. headed
shot) in both regressions used to compute rGAX. At the same time, if the sole
interest is in identifying the ability to shoot with the foot (or head), then
only these types of shots should be taken into account. Although this may reduce
sample size drastically, this effect is captured in the uncertainty
bounds provided by rGAX. Finally, by explicitly modeling the propensity of a
player taking a shot given the circumstances of the shot, i.e., by regressing
$X^p$ on $Z$, rGAX account for the fact that skilled players possess a shooting
profile that differs from that of the average player. Hence, rGAX address (3)
and are shown to be more robust to the data used for computing the xG model. Our
results show that, in contrast to GAX, rGAX are less affected when estimating an
xG model from data containing an overrepresentation of a certain set of players.
Lastly, although we cannot directly address the instability issue due to lack
of data for more seasons, rGAX offers new perspectives for evaluating stability
by using $p$-values and confidence intervals, which we leave for future work.

In summary, we demonstrate that rGAX and, in general, residualized metrics of a
similar form (see Table~\ref{tab:applications} and Appendix~\ref{app:use_cases})
provide a step toward more effectively measuring player skills. In the
following, we discuss relevant practical considerations when using these
residualized metrics for player evaluation. 

\subsection{Practical considerations}\label{subsec:pc}

We end with practical issues that may arise when employing the semiparametric
approach outlined in Section~\ref{sec:PLLM}. In particular, we address (i) the
independence assumption that makes inference possible using the described
methodology, (ii) the issue of small effective sample sizes due to limited goals
in soccer, (iii) the need for multiplicity corrections, (iv) testing practically
relevant differences via non-nil null hypotheses, and (v) the choice of control
variables for the regression models. While we focus on shooting skill evaluation
in soccer in this section, these considerations are relevant for other
disciplines as well.

\paragraph{Independence assumption.} The validity of $p$-values and confidence
intervals derived based on our approach is conditional on having independent
observations. This assumption may be challenged due to the sequential nature of
soccer games. For instance, the success of a rebound shot may depend on the
prior shot that led to the rebound. However, when the set of conditioning
variables $Z$ is sufficiently rich and the null hypothesis is assumed to be true
(i.e., player $X^p_j$ is irrelevant for the prediction of $Y$ given $Z$), it may
still be reasonable to assume that shots are independent conditional on the
circumstances under which a shot has taken place, as described by $Z$.

\paragraph{Effective sample size.} As noted in Section~\ref{sec:Intro}, goals in
soccer are rare, with only 10\% of shots being converted into a goal. Therefore,
care needs to be taken when tuning the outcome regression $\Ex[Y \given Z]$. A
similar argument holds for players that rarely occur in a dataset. In our
experiments, we circumvented this issue by considering only players who shot at
least 20 times and scored at least one goal among these shots.

\paragraph{Multiplicity corrections.} When testing several null hypotheses, the
family-wise error rate (FWER, i.e., the risk of at least one type~I error) and
false discovery rate (FDR, i.e., the proportion of false rejections among all
rejections) increase. Therefore, if several players are evaluated and the
results are used for decision making (e.g., player transfers), the resulting
$p$-values ought to be corrected \citep{bender2001}. For controlling the FWER, a
Bonferroni-Holm \citep{holm1979simple} adjustment can be used, while for
controlling the FDR, a Benjamini-Hochberg correction can be used
\citep{benjamini1995}. The former makes no assumption on the dependence between
$p$-values, while the latter is valid under certain kinds of positive
dependence. If the assumption of independent or positively dependent $p$-values
is not tenable, the Benjamini–Yekutieli procedure can be applied.

\paragraph{Non-nil null hypotheses.} A common criticism of null hypothesis
testing is that the rejection of a null hypothesis does not imply a practically
relevant effect size \citep{altman1995}. While this criticism is valid for
so-called ``nil null hypotheses'' (such as $H_0: \operatorname{GAX} = 0$),
non-nil null hypotheses (such as $H_0 : \operatorname{GAX} >
\operatorname{GAX}_0$ where $\operatorname{GAX}_0 > 0$ denotes the smallest
effect size of interest) circumvent this issue \citep{nunnally1960}. By
Proposition~\ref{thm:prp1}, our framework already allows for directional tests
within the PLLM. Our approach also allows the specification of such a minimal
relevant effect on the scale of the expected conditional covariance between the
player indicator and the outcome (such as GAX).

\paragraph{Controls for regression models.} Deciding on the features to include
in $Z$ is not only critical to ensure conditional independence between
shots, but is also important to accurately identify shooting skill. In
particular, it is important to identify ``good'' and ``bad'' control variables
\citep{Cinelli24controls}, i.e., features needed to identify shooting skills and
features that may obscure shooting skills. A particular example is the inclusion
of team strengths. While including defensive team strength to accurately model
xG values is sensible, the inclusion of offensive team strengths should be
avoided. This is due to the fact that a player has substantial influence on a
team's strength, or in other words, a team's strength is dependent on a player's
skill. For a complete list of features included and omitted for computing GAX
and rGAX, see Appendix \ref{app:data} and our GitHub repository at
\url{https://github.com/Rob2208/rGAX_and_beyond}. 
In light of this, it is imperative to select the features comprising
$Z$ for each application carefully. In our case, we derive a different set of
features $Z$ for each of our use cases presented in Appendix
\ref{app:use_cases}.

\subsection*{Acknowledgments} This work was supported by the WU Project 2025
``Learning causal structures from dependent data'' from the  Anniversary
Foundation of the Vienna University of Economics and Business. We thank Bettina
Gr\"un and Lukas Sablica for fruitful discussions.

\bibliographystyle{abbrvnat}
\bibliography{References}

\begin{appendix}

\section{Additional results for rGAX}
\label{app:rGAX_add}

\begin{figure}[t!]
    \centering
    \includegraphics[width=0.95\textwidth]{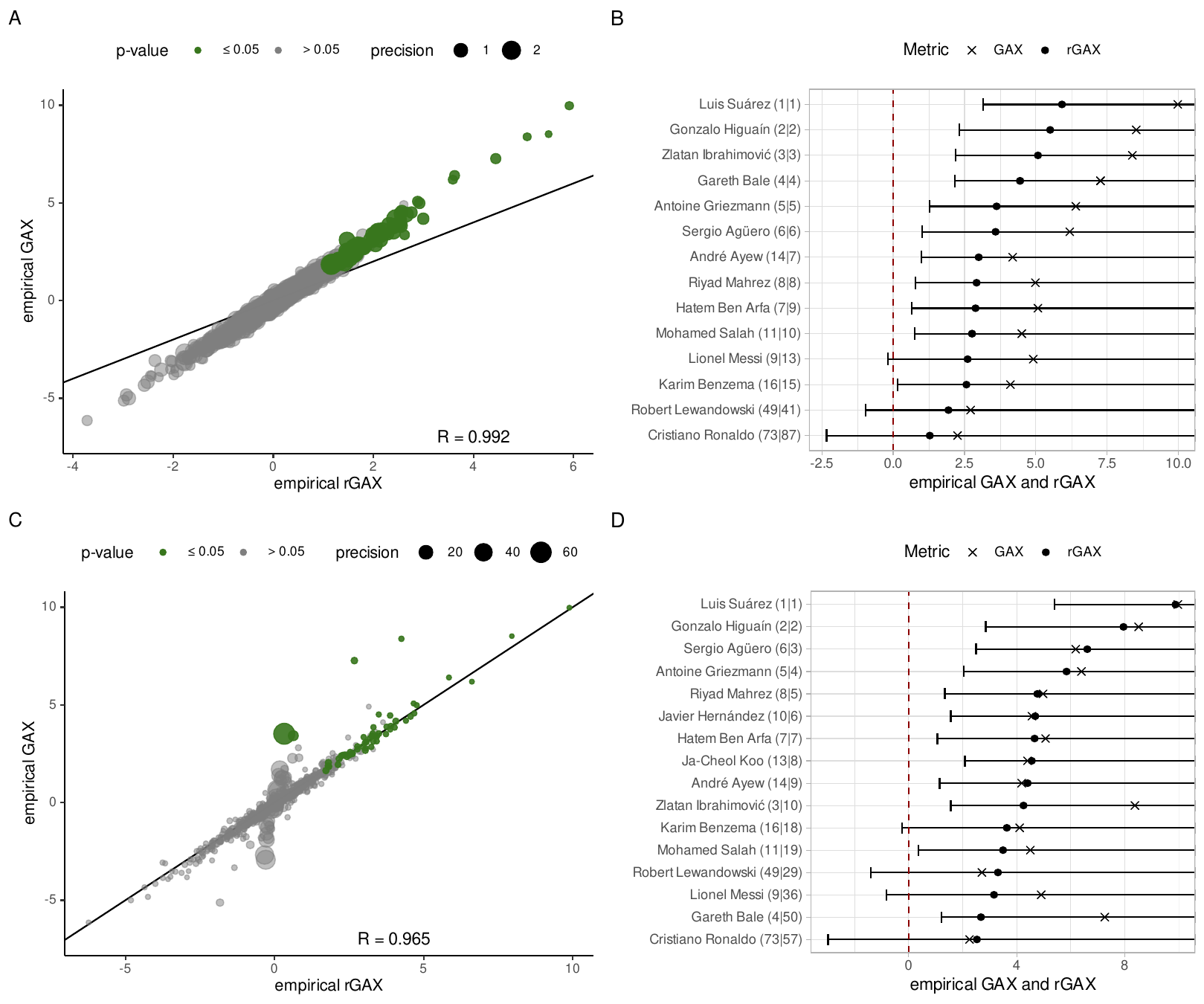}
    \caption{
    Goals and residualized goals above expectation plots for the 2015/16 season
    of the 5 big European soccer leagues using two different models for the
    regression of $X^p$ on $Z$. A: Scatterplot of empirical GAX and rGAX using an
    untuned random forest. The solid line indicates the identity. The
    correlation coefficient $R$ is added to the plot. B: Player-wise empirical
    GAX and rGAX from an untuned random forest with one-sided 95\% confidence
    intervals for rGAX (without multiplicity corrections) for the top 10
    players with respect to empirical rGAX and 5 selected well-known players.
    C and D: Similar to A and B, using a tuned xgboost for the regression of
    $X^p$
    on $Z$.
    }
    \label{fig:xreg_over_underfit}
\end{figure}

In this section, we discuss the choice of machine learning models for the
regressions of $X^p$ , $p = 1, \ldots, P$, on $Z$. While we rely
on recent literature on xG models for
the regression of $Y$ on $Z$, the regression of $X^p$ (the player
indicator) on
$Z$ has not yet been studied in the sports data science literature. A
particularly delicate point is the small effective sample size of players
shooting the ball, and hence the high imbalance in the outcome. In the main
text, we used an out-of-bag tuned random forest for the regression, as random
forests are well known for their flexibility and competitive performance under
little amount of tuning
\citep{Breiman2001RF,FernandezDelgado2014RF,Dandl24forests}. 

\begin{figure}[t!]
    \centering
    \includegraphics[width=0.95\textwidth]{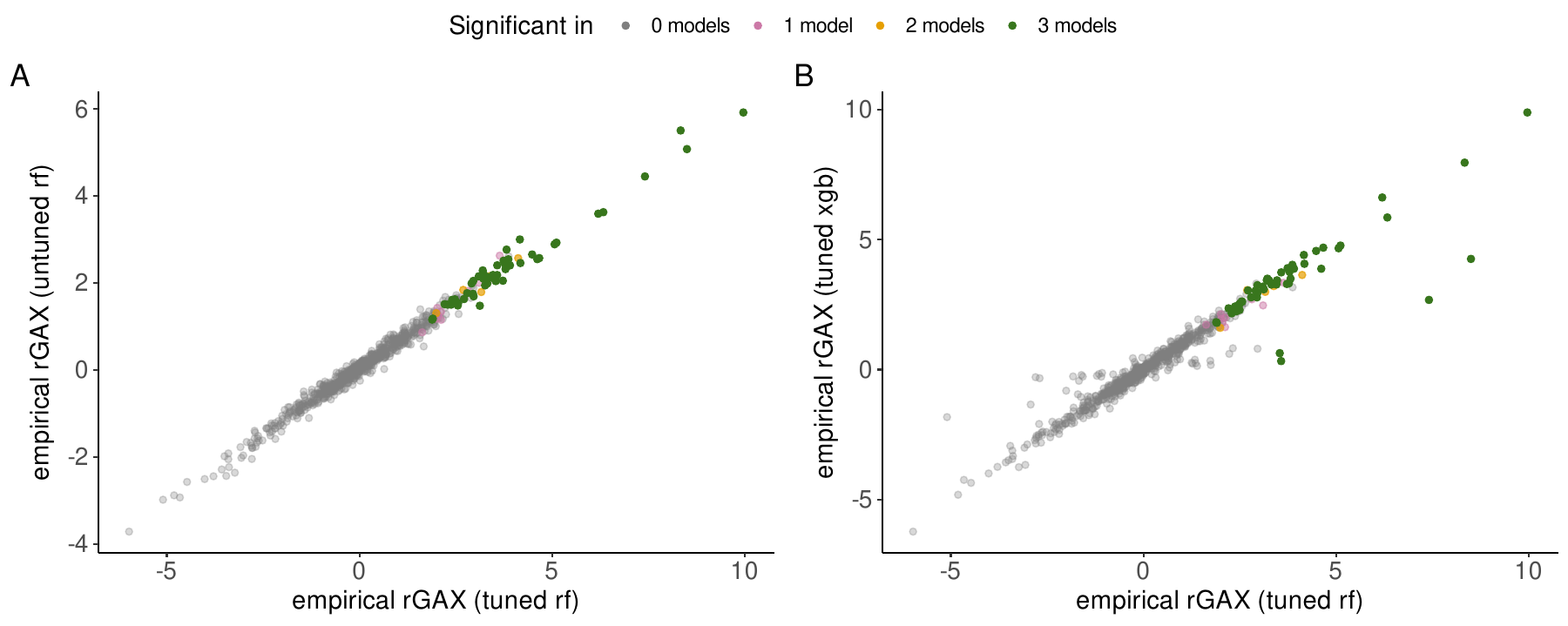}
    \caption{
    Scatterplot of empirical rGAX values as obtained from different models for the
    regression of $X^p$ on $Z$. A: Scatterplot of empirical rGAX from an 
    untuned random forest and empirical rGAX from a tuned random forest. 
    B: Scatterplot of empirical rGAX from a tuned xgboost model and empirical
    rGAX from a tuned random forest.
    }
    \label{fig:xreg_comp}
\end{figure}

Figure~\ref{fig:xreg_over_underfit} displays the results when using a simple
random forest (A and B) without tuning and a thoroughly tuned xgboost (C and D)
model respectively. For the random forest without tuning, the regression
underfits the data, and we observe that the empirical rGAX values are shifted
towards being closer to zero in contrast to empirical GAX. For the tuned xgboost
regression, the general pattern is captured quite well, but we observe outliers
due to overfitting to the data. While these results suggest that the regression
method for $X^p$ may have an impact on the final rGAX estimate, the
double robustness property of the GCM test statistic (see
Section~\ref{sec:PLLM}) ensures that the inference remains valid even when the
two regressions converge at a slower rate. Figure~\ref{fig:xreg_comp} displays a
scatterplot of the empirical rGAX values as obtained from the untuned random
forest (A) and the tuned xgboost (B), respectively, against the rGAX values from
the tuned random forest used in the main text. While there are subtle
differences in the estimates, we highlight the corresponding significant values
(at the 5\% level) in each model. Most of the significant players are shared
among all three models, emphasizing the robustness of rGAX.

\section{Further use cases of the proposed framework}
\label{app:use_cases} 

\subsection{Basketball: Quantified shooter impact}
\label{subsec:SQ_SI}

A common approach to measure shooting skills in basketball is to calculate a
player's field goal percentage (FG\%, \citealp{DalyGrafstein19BBeFG}). However,
similar to counting goals in soccer, FG\% simply averages the number of shots
taken by a player. Hence, it does not account for the circumstances in which a
shot was taken. In order to accurately capture a player's shooting skill,
researchers have developed xG models for basketball
\citep{Chang14BBxG,DalyGrafstein19BBeFG,Metulini20SI}. Similar to soccer, the xG
model in basketball is more commonly termed shot quality (SQ) model
\citep{Chang14BBxG} and estimates the probability of scoring from a shot
depending on shot-specific features. 

In order to determine the shooting skill of a player, \cite{Metulini20SI} use
their SQ model and compute the average difference between the actual outcome and
the model's predictions. This can be seen as an analogue to GAX in soccer,
albeit they additionally divide by the number of attempts a player took. To
account for the fact that there are different types of shots (two-point and
three-point shots), \cite{Metulini20SI} compute a shooting quality value for
each type of shot, i.e., obtaining one value for two-point shots and one value
for three-point shots. In contrast, \cite{Chang14BBxG} model shooter quality by
directly accounting for the different point values in shot types. In particular,
they consider the average difference between a weighted outcome (in basketball
commonly known as effective field goal percentage, eFG\%) and a weighted SQ
value, where three-point shots obtain a weight of 1.5.

More formally, given i.i.d.\ shot data $\{(Y_j, X^p_j, Z_j)\}_{j=1}^N$,
 empirical quantified shooter impact (qSI) of player $p$ for measuring 
shooting skill in basketball can be computed as 
\begin{equation}\label{eq:player_value_bb} 
\widehat{\operatorname{qSI}}_p \coloneqq \sum_{j=1}^{N} (Y_j - \hat h(Z_j))X^p_j,
\end{equation}
where $Y$ is again the outcome of a shot, $X^p$ is a player indicator, and
 $\hat
h$ is an estimator for $h(Z) = \Ex[Y \given Z]$, the conditional
expectation of a  shot
 given features $Z$. Using the approach from
\cite{Metulini20SI}, $Y$ 
is a binary outcome variable indicating success (1) or failure (0) of a shot.
Hence, in this case, qSI is completely analogous to GAX. Following the approach
of \cite{Chang14BBxG}, $Y$ can also be the outcome of a shot (0 if the shot 
was missed, 2 for two-point shots, and 3 for three-point shots). Using our
findings in Section~\ref{sec:PLLM}, we propose to measure shooting skill of player $p$ in
basketball via empirical residualized qSI (rqSI)
\begin{equation}\label{eq:rsQI} 
\widehat{\operatorname{rqSI}}_p \coloneqq \sum_{j=1}^{N} (Y_j - \hat
h(Z_j))(X^p_j- \hat f_p(Z_j)),
\end{equation}
where $\hat f_p(Z)$ is an estimator of $f_p(Z) =
\Ex[X^p \given Z]$.
Independent of the outcome used, the GCM test again allows to obtain 
uncertainty quantification for rqSI in the form of $p$-values and confidence
intervals. Additionally, similar to rGAX, rqSI can again be interpreted as 
a player effect in a semiparametric model. Hence, a significant value implies
a player positively affecting the outcome of a shot.

\begin{figure}[t!]
    \centering
    \includegraphics[width=0.95\textwidth]{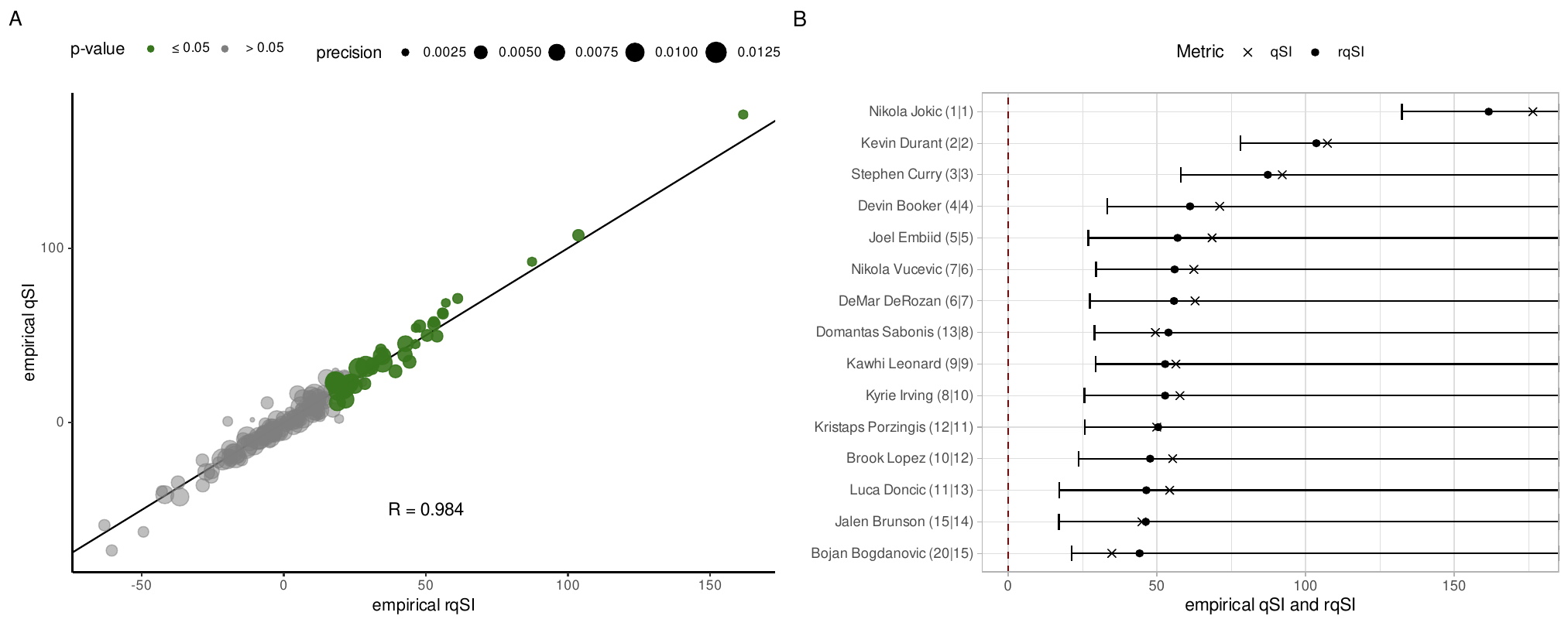}
    \caption{
    rqSI and qSI for the 2022/23 NBA season using a shot indicator as outcome
    (0 or 1).  A: Scatterplot of empirical qSI and rqSI. The size of
    the dots represents the estimated precision of the empirical
    rqSI. The solid line
    indicates the identity. The correlation coefficient $R$ is added to the
    plot. B: Player-wise empirical qSI and rqSI with one-sided 95\% confidence
    interval for rqSI (without multiplicity corrections) for the top 15
    players with respect to empirical rqSI. The rank of the players in
    terms of qSI (left) and rqSI (right) is shown in brackets.
    }
    \label{fig:rqSI_indicator}
\end{figure}

\begin{figure}[t!]
    \centering
    \includegraphics[width=0.95\textwidth]{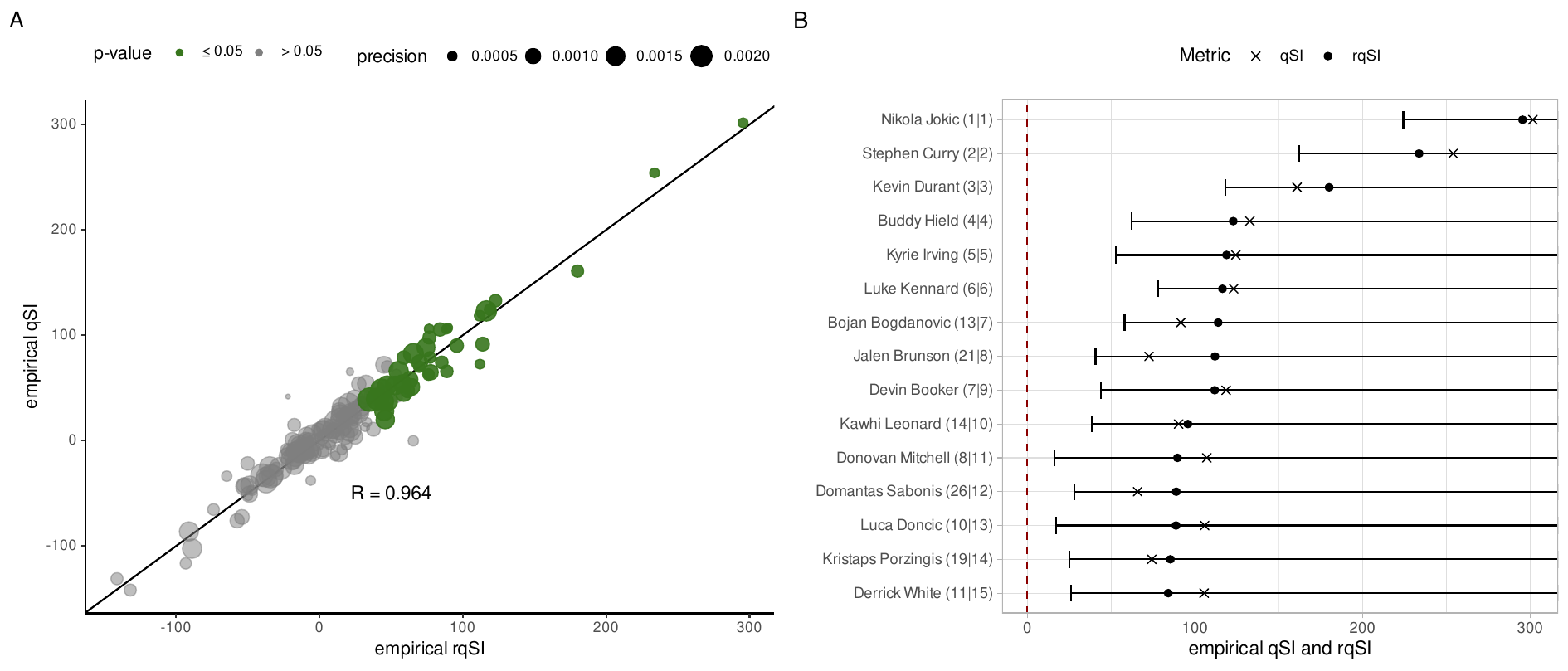}
    \caption{
    rqSI and qSI for the 2022/23 NBA season using the score value as outcome
    (0, 2, or 3).  A: Scatterplot of empirical qSI and rqSI. The size of
    the dots represents the estimated precision of the empirical rqSI. The
    solid line
    indicates the identity. The correlation coefficient $R$ is added to the
    plot. B: Player-wise empirical qSI and rqSI with one-sided 95\% confidence
    interval for rqSI (without multiplicity corrections) for the top 15
    players with respect to empirical rqSI. The rank of the players in
    terms of qSI (left) and rqSI (right) is shown in brackets.
    }
    \label{fig:rqSI_score}
\end{figure}

\begin{figure}[t!]
    \centering
    \includegraphics[width=0.95\textwidth]{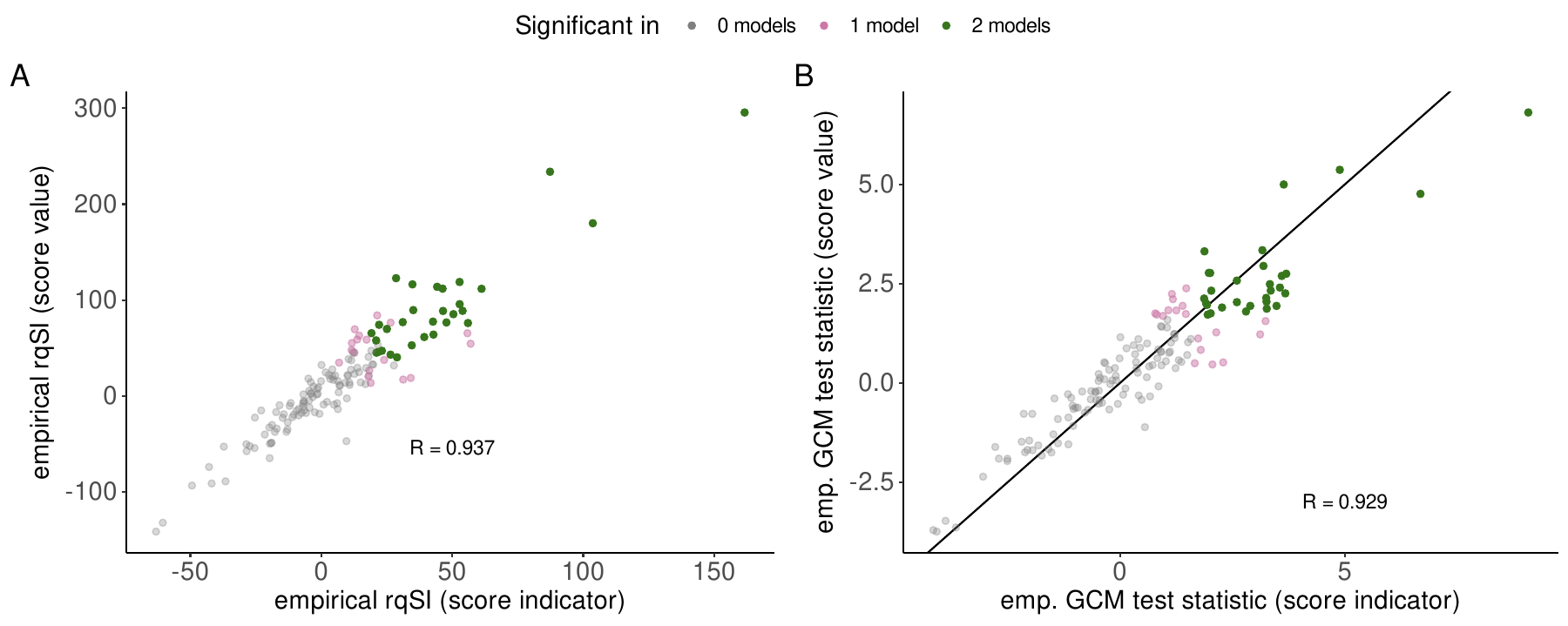}
    \caption{
    Comparison of rqSI values and GCM test statistic when using different
    outcomes.  A: Scatterplot of empirical rqSI when using score indicator
    outcome and score value outcome. B: Scatterplot of the empirical  GCM test
    statistic when  using score indicator outcome and score value outcome. The
    solid line in B indicates the identity. The correlation coefficient $R$ is
    added to both  plots.
    }
    \label{fig:rqSI_mod_comp}
\end{figure}

We apply our framework to data from the 2022/23 NBA season. We obtain
play-by-play data from this season via the \proglang{R} package \pkg{hoopR}
\citep{pkg:hoopR}. The package provides details on each shot, such as
$(x,y)$-coordinates and shot type, as well as contextual information such as
time played, score differential, and team and player information. We computed
empirical qSI and rqSI for all players with at least 300 shots in the 2022/23
season (including post-season), resulting in a total of 150 players. To obtain
empirical (r)qSI, $p$-values, and confidence intervals from the GCM test, the
\pkg{comets} package with an out-of-bag tuned random forest for both regressions
was used. Figure~\ref{fig:rqSI_indicator} shows the results for the binary
outcome of success or no success from shots. Figure~\ref{fig:rqSI_score} shows
the results using the actual score value as the outcome. In both cases, we
observe a high correlation between empirical qSI and rqSI. However, rqSI allows
for more insights by providing valid uncertainty quantification in the form of
$p$-values and confidence intervals. Furthermore, it allows for different
interpretation of the rqSI as a player effect in a semiparametric model on
either the probability of scoring (when using a binary outcome indicator) or the
scoring outcome (when using the score value as outcome). While there are slight
differences in the rankings when using different outcomes, Figure
\ref{fig:rqSI_mod_comp} shows that both models, in general, agree on a player's
shooting skill. In particular, the figure shows that the models mostly agree on
which players significantly impact the outcome of a shot (at the 5\% level).
While the scales of empirical rqSI values differ for the outcome types (Figure
\ref{fig:rqSI_mod_comp}A), the high correlation between empirical rqSI from
score indicator outcome and score value outcome demonstrates high agreement
between both approaches. Similarly, the GCM test statistics, i.e., the
standardized version of the empirical rqSI, which is directly related to the
strength parameter in our semiparametric framework, indicate strong agreement
between both approaches (Figure~\ref{fig:rqSI_mod_comp}B). 

\subsection{American football: Completion percentage above expectation}
\label{subsec:CPAE}

In American football, a popular metric to evaluate a quarterback's passing
skills is to calculate their completion percentage, i.e., the percentage of
passes that actually found a teammate. Similar to other metrics, simply 
counting the number of complete passes out of all pass attempts
does
not take into account the difficulty of a pass. Hence, the football analytics
community has developed pass completion probability (CP) models, taking into
account pass-specific features, which again may be seen as an analogue to an
expected goals model in soccer. Most prominently, the National Football League
(NFL) provides their own version of a CP model using player
tracking data via \href{https://nextgenstats.nfl.com/}{NFL Next Gen Stats}.
Aside from a number of CP models from various football analytics outlets (e.g.,
\href{https://www.pff.com/news/nfl-chicago-bears-justin-fields-most-accurate-quarteback-ohio-state-pff-college-era.}{PFF},
\href{https://fivethirtyeight.com/features/the-nfl-is-drafting-quarterbacks-all-wrong/}{FiveThirtyEight}),
the openly available \proglang{R}~package \pkg{nflfastR} \citep{pkg:nflfastR}
also provides a CP model.

To analyze passing skills of quarterbacks, a CP model can be used to calculate
completion percentage above expectation (CPAE, sometimes also called completion
percentage over expected, CPOE). Formally, CPAE can again be expressed using
\eqref{eq:player_value}. That is, given i.i.d pass data $\{(Y_j, X^p_j,
Z_j)\}_{j=1}^N$,  the empirical CPAE of player $p$ can be written as 
\begin{equation}\label{eq:player_value_nfl} 
\widehat{\operatorname{CPAE}}_p \coloneqq \sum_{j=1}^{N} (Y_j - \hat h(Z_j))X^p_j,
\end{equation}
where $Y$ is the outcome of a pass (1: complete, or 0: incomplete), $X^p$ is  a
player indicator, and
$\hat h(Z)$ is an estimator for $h(Z) = \Ex[Y \given Z]$, the
conditional expectation of a pass  given features

$Z$. Hence, $\hat h$ represents a
completion probability model fitted to the data. Analogous to rGAX, we propose
to instead use residualized CPAE (rCPAE) to measure passing skill of a
quarterback $p$, with the empirical version defined as
\begin{equation}\label{eq:rCPAE} 
\widehat{\operatorname{rCPAE}}_p \coloneqq \sum_{j=1}^{N} (Y_j - \hat
h(Z_j))(X^p_j- \hat f_p(Z_j)),
\end{equation}
with $\hat f_p$ being an estimator of $f_p(Z) =
\Ex[X^p \given Z]$. rCPAE 
again allows for valid uncertainty quantification via the GCM test, and
can conveniently be interpreted as a player's effect on the completion probability
in a semiparametric model of the form of \eqref{eq:pllm_lo}.

\begin{figure}[t!]
    \centering
    \includegraphics[width=0.95\textwidth]{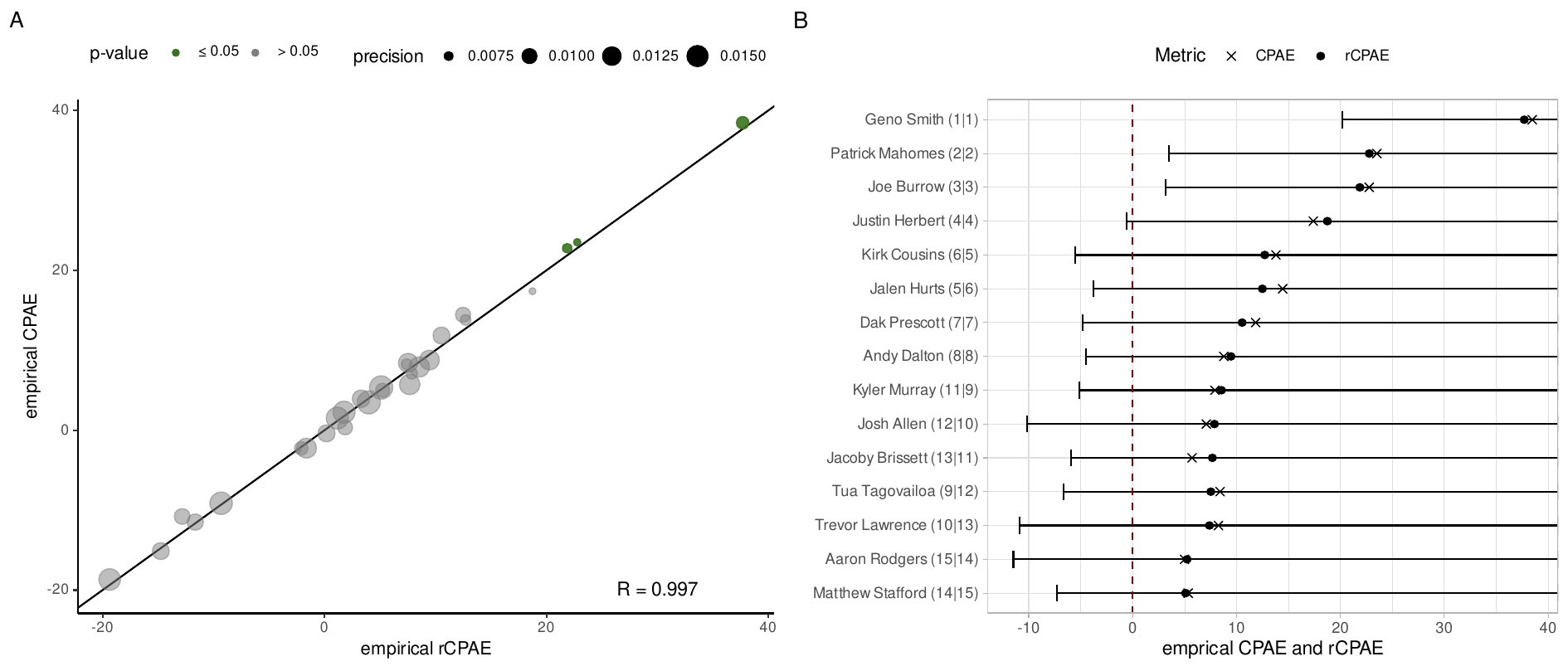}
    \caption{
    rCPAE and CPAE for the 2022/23 NFL season.  A: Scatterplot of empirical
    rCPAE and CPAE. The size of the dots represents the estimated
    precision of the empirical rCPAE. The solid line
    indicates the identity. The correlation
    coefficient $R$ is added to the plot. B: Player-wise empirical rCPAE and
    CPAE with one-sided 95\% confidence interval for rCPAE (without multiplicity
    corrections) for the top 15 players with respect to empirical rCPAE.
    The rank of the players in terms of CPAE (left) and rCPAE (right) is
    shown in brackets.
    }
    \label{fig:rCPAE}
\end{figure}

We compute empirical CPAE and rCPAE values for all quarterbacks with at least
300 passing attempts in the 2022/23 NFL season. To do so, we obtain NFL
play-by-play data from the \pkg{nflfastR} package. All results are again
obtained via the \pkg{comets} package with an out-of-bag tuned random forest for the
regressions. Figure~\ref{fig:rCPAE} shows the result for evaluating passing
skills of quarterbacks. The scatterplot in Figure~\ref{fig:rCPAE}A shows a high
correlation between empirical CPAE and rCPAE. Figure~\ref{fig:rCPAE}B highlights
the advantages of rCPAE, showing the top 15 quarterbacks with respect to their
empirical rCPAE as well as the 95\% one-sided confidence intervals for the players.

\subsection{Soccer: Time to first injury}
\label{subsec:TTI}

We consider extensions of our residualized player evaluation metrics to settings
with time-to-event responses, in particular, to the injury-proneness of players
in soccer.
In time-to-event analysis, the response is a positive real valued random
variable $Y^* \in \RR_+$, which indicates the time a certain event has happened.
In our case, this event corresponds to an injury of a player.
This response is modeled in terms of features $(X^p, Z) \in
\mathcal{X} \times \mathcal{Z}$, where $X^p$ denotes 
the indicator for player $p$
and $Z$ denotes other features. For brevity, we denote ${W^p} \coloneqq
(X^p, Z)$ and
$\mathcal{W} \coloneqq
\mathcal{X} \times \mathcal{Z}$. In practice, this event time may only be
partially observed due to loss of follow-up or competing events. Instead, we
observe a \emph{censored} version of $Y^*$, which is the event time or a
censoring time $C \in \RR_+$,
\[
Y \coloneqq \min(Y^*, C).
\]
Besides $Y$, we also observe the indicator random variable
$\delta \coloneqq \1(Y^* \leq C)$, which indicates whether an observation
corresponds to an event ($\delta = 1$) or was censored ($\delta = 0$).

The target in survival analysis is an estimate of the (cumulative) hazard of an
event happening before a given time. We denote the cumulative hazard function by
${\Lambda_p} : \RR_+ \times \mathcal{W} \to \RR_+$ and a closed-form
expression for
${\Lambda_p}$ is given by ${\Lambda_p}(y, w) = -\log(1 -
F_{Y\given {W^p}}(y \given w))$,
where $F_{Y \given {W^p}}(y \given w)$ denotes the conditional cumulative
distribution function of $Y$ given ${W^p}$ \citep{klein2014handbook}.
Further, the
martingale residual \citep{therneau1990} for an observation $(y, \delta, w)$ is
given by
\[
\delta - {\Lambda_p}(y, w).
\]
Martingale residuals can be interpreted as the difference between the observed
and expected number of events up to time point $y$ given the features $w$.

We now consider testing the null hypothesis $H_0^* : Y^* \indep X^p
\given Z$, which involves the true but unobserved event time $Y^*$. We assume
that $C \indep Y^* \given W^{p}$ (\emph{uninformative} censoring) for
the validity of the involved survival regression methods and the conditional
independence test, introduced below.

Given i.i.d.\ observations $\{(Y_j, \delta_j, X^p_j, Z_j)\}_{i=1}^N$ 
and an estimator
$\hat\Lambda$ of the cumulative hazard function under $H_0$ (which,
therefore, no longer depends on $p$), we define
the empirical injuries above expectation (IAX)
\[
\hat{\operatorname{IAX}}_p \coloneqq \sum_{j=1}^n (\delta_j - \hat\Lambda(Y_j,
Z_j)) X^p_j,
\]
which, analogously to the GAX case, correspond to the unscaled score statistic
in a partially linear Cox model. However, the score test derived from a
partially linear Cox model does not yield valid inference without imposing
strong parametric regularity conditions.

The TRAM-GCM test, as introduced in \citet{kook2024tramgcm}, generalizes the GCM
test to censored responses and corresponds to what we define as empirical
residualized IAX (rIAX),
\[
\hat{\operatorname{rIAX}}_p \coloneqq \sum_{j=1}^N (\delta_j -
\hat\Lambda(Y_j, Z_j)) (X^p_j - {\hat{f}_p}(Z_j)),
\]
where ${\hat{f}_p}$ is an estimator of $f_p(Z) \coloneqq
\Ex[X^p \given Z]$. Under the
assumptions outlined in Theorem~15 \citep{kook2024tramgcm}, the test based on
rIAX enjoys the same double robustness properties as the GCM test \citep{SP20}.
\citet[Proposition~22 in Appendix~A]{kook2024tramgcm} also show that the
TRAM-GCM test corresponds to testing $H_0: \beta = 0$ under a partially linear
Cox model, defined by $\log \Lambda(y, x, z) = \log \Lambda_0(y) + x\beta +
g(z)$, where $\Lambda_0$ denotes the baseline cumulative hazard and $g$ is an
arbitrary measurable function. 
In case of longitudinal survival
data and time-constant features, the TRAM-GCM test for the Cox model coincides
with the endpoint local covariance measure test introduced in
\citet{christgau2023nonparametric}. 

\begin{figure}[t!]
    \centering
    \includegraphics[width=\linewidth]{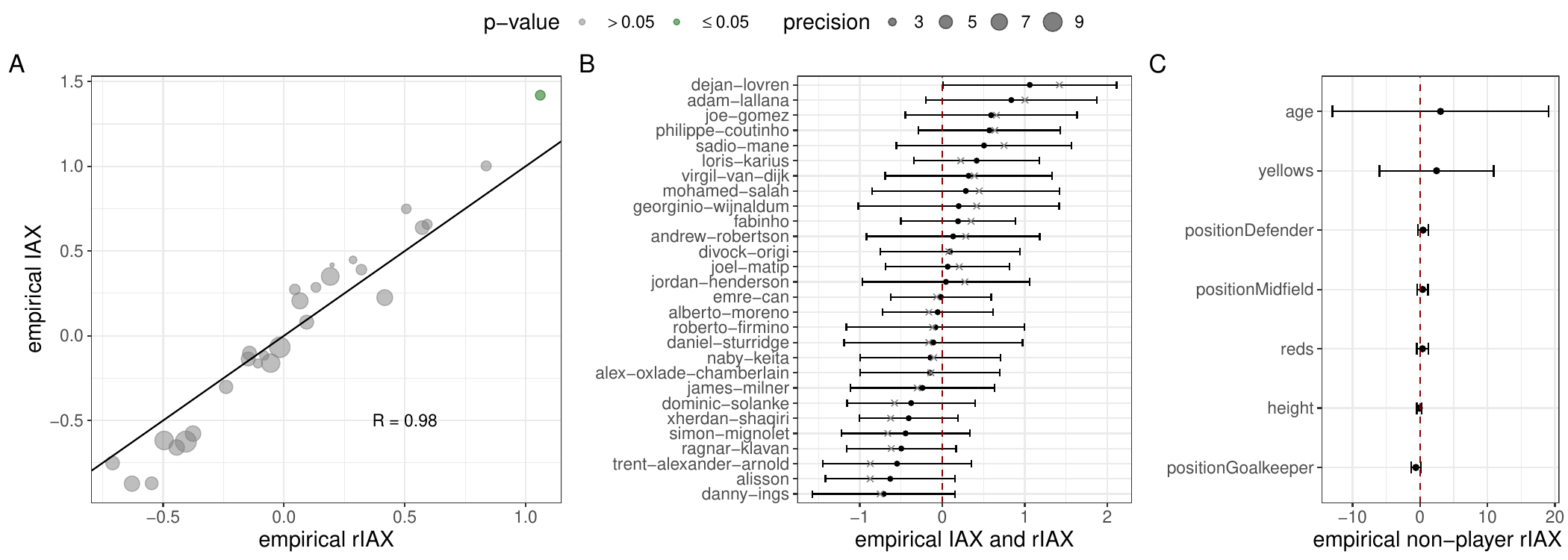}
    \caption{%
    Events and residualized events above expectation plots for time to first
    injury in the Liverpool F.C.\ data. A: Scatterplot of  the empirical IAX and
    rIAX. The size of the dots represents the estimated precision of 
    the empirical rIAX. 
    The solid line indicates the identity. 
    The correlation coefficient $R$ is added to the plot.
    B: Player-wise  empirical IAX
    and rIAX with 95\% confidence intervals for rIAX (without multiplicity
    corrections). C: Feature-specific rIAX with 95\% confidence intervals
    (without multiplicity corrections). For the survival regression, a
    random survival forest was used. For the feature residualization step, a
    random forest was used. The vertical dashed line in B and C indicates the
    null hypothesis of no effect.
    }
    \label{fig:reax}
\end{figure}

\paragraph{Example application: Time to first injury in soccer} We apply the
proposed player evaluation framework to evaluate injury-proneness of soccer
players in the Liverpool football club for the seasons 2017/2018 and 2018/2019.
The data are openly available through the \pkg{injurytools} 
\citep{pkg:injurytools}
\proglang{R}~package and contain information on 28 players. For this
illustration, we consider time to first injury within each season and treat the
injuries (and thereby players) as independent conditional on their age, height,
number of yellow or red cards and position and the season. In total, the data
contains 42 rows, of which 33 correspond to events and 9 to right-censored event
times. We use a random survival forest model to obtain $\hat\Lambda$ and a
random forest to obtain $\hat{f}_p$. Figure~\ref{fig:reax} shows the results:
Figure~\ref{fig:reax}A shows a strong correlation between IAX and rIAX.
Figure~\ref{fig:reax}B shows the player-specific IAX and rIAX, together with
95\% confidence intervals for the latter. rIAX, for most players, is closer to
zero than IAX. Figure~\ref{fig:reax}C, in addition, shows rIAX with 95\%
confidence intervals for the hypotheses that time to first injury is independent
of a given feature conditional all other features (and season). We refer
to (empirical) rIAX that is not computed with $X^p$ being a player
indicator as (empirical) non-player rIAX. A large positive value of rIAX can be
interpreted as a high proneness for early injury. However, even without
adjusting for multiple testing and, likely, the small size of the dataset, there
is no substantial evidence for any of the players to be more injury prone.

\section{Data}\label{app:data}

Table~\ref{tab:features} provides a detailed list of the shot-specific features
extracted from the event stream data and used for xG models in
Section~\ref{sec:res}. Additionally, as mentioned in the main text (see
Section~\ref{sec:data}), we compute team strength parameters for the offensive
and defensive team for each shot (for details on the derivation, see
Appendix~\ref{app:comp}). For (r)GAX, we purposefully omit the offensive team
strength parameter as a feature, since we consider it a bad control variable
(see Section \ref{sec:disc}). Similarly, for (r)GSAX, the defensive team
strength is considered a bad control and hence omitted from the feature set.

\begin{table}
\caption{\label{tab:features}Features engineered from data.}
\resizebox{\textwidth}{!}{
\begin{tabular}{llp{10cm}}
\toprule
\bf Variable & \bf Type & \bf Description \\
\midrule
shot.type.name & categorical & Shot type (one of Open Play, Corner, Free
Kick) \\
shot.technique.name & categorical & Shot technique (one of Normal,
(Half) Volley, Diving Header, Backheel, Lob, Overhead Kick) \\
shot.body\_part.name & categorical & Shot body part (one of Head, Foot,
Other) \\
DistToGoal & numeric & Distance to center of the goal \\
DistToKeeper & numeric & Distance of goalkeeper to goal \\
DistSGK & numeric & Distance to goalkeeper \\
distance.ToD1 & numeric & Distance to closest defender in front of
goal \\
distance.ToD2 & numeric & Distance to 2nd closest defender in front of
goal \\
distance.ToD1.360 & numeric & Distance to closest defender (general) \\
distance.ToD2.360 & numeric & Distance to 2nd closest defender
(general) \\
AngleToGoal & numeric & Angle between shot and the center of the goal
(in degree) \\
AngleToKeeper & numeric & Angle between goalkeeper and center of goal
(in degree) \\
AngleDeviation & numeric & Absolute difference in the two angles \\
angle & numeric & Angle between shot and goal posts (in radians) \\
AttackersBehindBall & integer & Attackers behind the ball (in $x$
coordinate) \\
DefendersInCone & integer & Defenders in cone drawn from shot to goal
posts \\
DefendersBehindBall & integer & Defenders behind ball (in $x$
coordinate) \\
density & numeric & Free space for shooter, sum over the inverse of
distances from shooter to defenders \\
density.incone & numeric & Sum over the inverse of distances from
shooter to defenders in cone \\
\bottomrule
\end{tabular}}
\end{table}

\section{Computational details}\label{app:comp}

\subsection{Hyperparameter tuning}
\label{subsec:hyperpar}

We used two main types of models (xgboost and random forest) for the two
regressions involved in the GCM test and to obtain residualized metrics (rGAX,
rGSAX, rqSI, rCPAE, and rIAX). We briefly describe the hyperparameter tuning for
both.

For the xG (psxG) models used for the regression of $Y$ on $Z$ to obtain
empirical rGAX (rGSAX), we used a gradient boosted tree ensemble method
implemented in the \pkg{xgboost} package. To choose hyperparameters, we set up a
cross validation routine on a grid of values for the learning rate \texttt{eta}
and the parameter \texttt{max\_depth}, i.e.,~the maximum depth of the trees used
for the tree ensemble. For \texttt{eta}, we consider values in
$\{0.001,0.005,0.01,0.1,0.5,1\}$, for \texttt{max\_depth} values in
$\{1,3,4,5,7,9\}$. Additionally, we perform early stopping to determine the
optimal number of boosting iterations. All other regressions in this paper were
fitted using a tuned (survival) random forest. Instead of using cross validation
to determine the optimal hyperparameters, random forest can be conveniently
tuned using out-of-bag (OOB) data. In particular, random forests are estimated
using trees fitted to bootstrap samples of the data and random subsamples of the
features. Thereby, not all data is used for every single tree. The data not used
is termed OOB data, and model performance can be evaluated on the OOB data. In
particular, we again use a grid of values for the tuning parameters
\texttt{mtry} (number of randomly selected candidate variables to split on in
each tree), and \texttt{max.depth} (the maximum depths of the trees) and select
the optimal set of parameters using the OOB error. For \texttt{mtry}, we
consider values in $\{1,\sqrt{p},p\}$, where $p$ is the number of features used
for the regressions (i.e., the dimension of $Z$), for \texttt{max.depth} values
in $\{1,\dots,5\}$.

\subsection{Team strength estimates}
\label{subsec:ts}

We describe the usage of Poisson generalized linear models to obtain
offensive and defensive team strengths, which were used as features in the
regressions for computing rGAX in Section~\ref{sec:res}. We follow the approach
in \citet{KN03}, according to which the bivariate Poisson model can be
formalized in the following way.
For $M$ matches featuring a total of $T$ teams, we write $Y_{i j m}$ the random
variable number of goals scored by team $i$ against team $j$ $(i, j \in\{1,
\ldots, T\})$ in match $m$ (where $m \in\{1, \ldots, M\}$ ). The joint
probability function of the home and away score is then given by the bivariate
Poisson probability mass function,
\begin{equation}
\label{eq:biv_po_lh}
\begin{aligned}
\mathbb{P}\left(Y_{i j m}=z, Y_{j i m}=y\right)= & \frac{\lambda_{i j m}^z
\lambda_{j i m}^y}{z ! y !} \exp \left(-\left(\lambda_{i j m}+\lambda_{j i
m}+\lambda_C\right)\right) \ \cdot \\
& \sum_{k=0}^{\min (z, y)}
\binom{z}{k}
\binom{y}{k}
k !\left(\frac{\lambda_C}{\lambda_{i j m} \lambda_{j i m}}\right)^k,
\end{aligned}
\end{equation}
where $\lambda_C$ is a covariance parameter assumed to be constant over all 
matches and $\lambda_{i j m}$ is the expected number of goals for team $i$
against team $j$ in match $m$, which are modeled as
\begin{equation}
\label{eq:po_exp}
\log \left(\lambda_{i j
m}\right)=\beta_0+\left(\text{att}_i-\text{def}_j\right)+h \cdot
\mathds{1}(\text {team } i \text { playing at home}) \text {, }
\end{equation}
where $\beta_0$ is a common intercept and $\text{att}_i$ and 
$\text{def}_j$ are the attacking and defensive strength
parameters of teams $i$ and $j$, respectively. 
Since the attacking and defensive parameters are unique
up to addition by a constant, the constraint that the sum of these
has to equal zero is used. The last term $h$ represents the home effect and is
only added if team $i$ plays at home. Note that the bivariate Poisson model
corresponds to an independent Poisson model if $\lambda_C=0$.

Using historic match data obtained from \url{https://www.football-data.co.uk/}, 
we estimate the strength parameters via maximum likelihood estimation. To 
account for the fact that team strengths vary in time, and we are mostly
interested in the actual strength we use a weighted maximum likelihood approach, 
i.e., we maximize the weighted log-likelihood function
\begin{align}
\label{eq:wll}
\ell(\theta \given y_{im},y_{jm}) = \sum_{m = 1}^M w_m \log(\Prob(Y_{im} = y_{im},Y_{jm} = y_{jm}\given \theta)),
\end{align}
where $\theta = (\beta_0,\text{att}_1,\dots,\text{att}_T,\text{def}_1,\dots,\text{def}_T,h,\lambda_C)$ 
is the set of all parameters to be estimated, and $w_m$ is a 
weight accounting for the recency of the match. We follow existing
literature and set it to $w_m = (\frac{1}{2})^{\frac{d}{p}}$, where $d$
represents the number of days passed since the match, and $p$ represents a relevant
period of interest. That is, a match played $p$ days ago contributes only half as
much as a match today \citep{GrollEA19}. Following the literature, we set $p =
500$. In this way, we obtain attacking and defensive strength parameters for
each team in our data set.

\section{Code example}\label{app:code}

We showcase how to obtain rGAX using the \pkg{comets} package for the case of Luis
Suárez. We first load correctly preprocessed data containing an indicator 
column for shots from Luis Suárez \\
(\texttt{Luis\_Suarez\_example.rds}).
\begin{verbatim}
R> library("tidyverse")
R> library("comets")
R> library("coin")
R> LS_data <- readRDS("Luis_Suarez_example.rds")
\end{verbatim}
To obtain rGAX, we use the GCM test implemented in the \pkg{comets} package.
\pkg{comets} allows us to define which machine learning model to use for the regressions
of $Y$ on $Z$ and $X$ on $Z$. As in the main text, we use a pre-trained xG model 
(loaded from the RDS-file \texttt{xg\_mod.rds}) for the
regression of $Y$ on $Z$ and a tuned random forest for the regression of
$X^p$ on $Z$.
For the former, we need to define a suitable regression method, while the latter is
pre-implemented in the package. To test whether $Y$ is independent of $X$ given $Z$,
the formula-based interface of the \texttt{comet} function can be used by providing
a formula of the type \texttt{Y \textasciitilde{} X | Z}. 
\begin{verbatim}
R> xg_mod <- readRDS("xg_mod.rds")
R> xG_reg <- function(y, x, xg_mod = NULL,...){
+   structure(xg_mod, class = c("xgb", class(xg_mod)))
+   }
R> set.seed(123)
R> GCM_suarez <- comet(shot_y ~ Luis_Suarez | . - Luis_Suarez, data = LS_data,
+    test = "gcm", reg_YonZ = "xG_reg", reg_XonZ = "tuned_rf",
+    args_YonZ = list(xg_mod = xg_mod), args_XonZ = list(probability = TRUE),
+    type = "scalar", verbose = 0, coin = TRUE)
\end{verbatim}
\begin{verbatim}
R> GCM_suarez

	Generalized covariance measure test

data:  comet(formula = shot_y ~ Luis_Suarez | . - Luis_Suarez, data = LS_data, 
    test = "gcm", reg_YonZ = "xG_reg", reg_XonZ = "tuned_rf", 
    args_YonZ = list(xg_mod = xg_mod), args_XonZ = list(probability = TRUE), 
    type = "scalar", verbose = 0, coin = TRUE)
Z = 3.5948, p-value = 0.0003247
alternative hypothesis: true E[cov(Y, X | Z)] is not equal to 0
\end{verbatim}
From the test result, we can extract $p$-values and the test statistic to see
that Luis Suárez has a significant positive impact on the probability of scoring
a goal in our semiparametric framework. With the GCM test result, we are also able
to obtain rGAX and the corresponding 95\% confidence interval. To compute 
the confidence interval, we use the \pkg{coin} package \citep{Hothorn08coin},
which relies on an approximation of the asymptotic permutation distribution
to estimate the standard deviation of the test statistic.
\begin{verbatim}
R> rGAX <- sum(GCM_suarez$rY * GCM_suarez$rX)
R> tst <- independence_test(GCM_suarez$rY ~ GCM_suarez$rX, teststat = "scalar")
R> sd <- sqrt(variance(tst))
R> ci <- c(rGAX - 1.96 * sd, rGAX + 1.96 * sd)
R> rGAX

[1] 9.969451

R> ci

[1]  4.533501 15.405400
\end{verbatim}

\section{Proof of Proposition~\ref{thm:prp1}}
\label{app:proofs}

\begin{proof}
We first prove $(i)$. We can write
\begin{equation}\label{eq-prp1-1}{
\begin{aligned}
\mathbb{E}[\operatorname{Cov}(Y,X^p \given  Z)] &= \mathbb{E}\big[\mathbb{E}[X^pY
\given  Z] - \mathbb{E}[Y \given  Z] \mathbb{E}[X^p \given  Z]\big] \\
&= \mathbb{E}\bigg[\mathbb{E}\big[X^p\mathbb{E}[Y \given  X^p,Z] \given  Z\big] -
\mathbb{E}[Y \given  Z] \mathbb{E}[ X^p \given  Z]\bigg],
\end{aligned}
}\end{equation}
where we have used the tower property of the conditional expectation in the
second equality. Since $X^p$ is binary,
we have that 
\[
\begin{aligned}
\mathbb{E}\bigg[\mathbb{E}\big[X^p\mathbb{E}[Y \given  X^p,Z] \given  Z\big]\bigg] &= 
\mathbb{E}\bigg[P(X^p = 1 \given  Z) \cdot 1 \cdot \mathbb{E}[Y \given  X^p = 1,Z]+P(X^p =
0\given  Z) \cdot 0 \cdot \mathbb{E}[Y \given  X^p = 0,Z]\bigg] \\
&= \mathbb{E}\big[f_p(Z)\mathbb{E}[Y \given  X^p = 1,Z]\big]
\end{aligned}
\] and \[
\begin{aligned}
\mathbb{E}\big[\mathbb{E}[Y \given  Z] \mathbb{E}[ X^p \given  Z]\big] &=
\mathbb{E}\big[\mathbb{E}[\mathbb{E}[Y \given  X^p,Z] \given  Z] f_p(Z)\big] \\
&= \mathbb{E}\big[f_p(Z) \big( f_p(Z) \mathbb{E}[Y \given  X^p = 1,Z] +
(1-f_p(Z))\mathbb{E}[Y \given  X^p = 0,Z] \big)\big] \\
&= \mathbb{E}\big[f_p(Z)^2 \big(\mathbb{E}[Y \given  X^p = 1,Z]-\mathbb{E}[Y \given  X^p =
0,Z]\big) +f_p(Z) \mathbb{E}[Y \given  X^p = 0,Z] \big]
\end{aligned}
\] 
Using these results, we obtain \[
\begin{aligned}
\mathbb{E}[\operatorname{Cov}(Y,X^p \given  Z)] &=
\mathbb{E}\big[\mathbb{E}\big[X^p\mathbb{E}[Y \given  X^p,Z] \given  Z\big]\big] -
\mathbb{E}\big[\mathbb{E}[Y \given  Z] \mathbb{E}[ X^p \given  Z]\big] \\
&= \mathbb{E}\big[\big(f_p(Z)-f_p(Z)^2\big) \big(\mathbb{E}[Y \given  X^p =
1,Z]-\mathbb{E}[Y \given  X^p = 0,Z]\big)\big]
\end{aligned}
\]
Since \(\big(f_p(Z)-f_p(Z)^2\big) > 0\) (as $0 < P(X^p = 1 \given  Z) < 1$ $P_Z$-almost surely),
$\mathbb{E}[\operatorname{Cov}(Y,X^p \given  Z)] = 0$ if and only if
$\big(\mathbb{E}[Y \given  X^p = 1,Z]-\mathbb{E}[Y \given  X^p = 0,Z]\big) = 0$.
Under the partially linear logistic model we have
\[
\mathbb{E}[Y \given  X^p,Z] = \frac{1}{1+e^{-g(Z)-X^p\beta}}.
\]
For $P_Z$-almost all $z$ and finite $g$ the function $x \mapsto
\frac{1}{1+e^{-g(z)-x\beta}}$ is a strictly monotone and thus injective function
if and only if $\beta \neq 0$. Therefore, the difference
$\big(\mathbb{E}[Y \given  X^p = 1,Z]-\mathbb{E}[Y \given  X^p = 0,Z]\big) = 0$ if and
only if \(\beta = 0\). Similarly, 
$\big(\mathbb{E}[Y \given  X^p = 1,Z]-\mathbb{E}[Y \given  X^p =
0,Z]\big) > 0 $ $(< 0)$ if and
only if $\beta > 0 $ $(< 0)$, hence $(ii)$ holds as well.  
This completes the proof of Proposition~\ref{thm:prp1}.
\end{proof}

\end{appendix}

\end{document}